\documentclass[longauth]{aa}
\usepackage{graphicx}
\usepackage{txfonts}
\usepackage{xspace}
\usepackage{xcolor}
\usepackage{url}
\usepackage[colorlinks=true, allcolors=blue]{hyperref}
\usepackage[flushleft]{threeparttable}
\usepackage{orcidlink}
\newcommand {\ixpe}{\text{IXPE}\xspace}
\newcommand{\smc}{\mbox{SMC~X-1}\xspace}
\newcommand{\lumcgs}{erg\,s$^{-1}$}

\begin{document}

\title{Probing the polarized emission from SMC~X-1: \\ the brightest X-ray pulsar observed by \ixpe}

\author{Sofia~V.~Forsblom \inst{\ref{in:UTU}}\orcidlink{0000-0001-9167-2790}
\and Sergey~S.~Tsygankov \inst{\ref{in:UTU}}\orcidlink{0000-0002-9679-0793}
\and Juri~Poutanen \inst{\ref{in:UTU}}\orcidlink{0000-0002-0983-0049}
\and Victor~Doroshenko\inst{\ref{in:Tub}}\orcidlink{0000-0001-8162-1105} 
\and Alexander~A.~Mushtukov  \inst{\ref{in:Oxford}}\orcidlink{0000-0003-2306-419X}  
\and  Mason Ng \inst{\ref{in:MIT}}\orcidlink{0000-0002-0940-6563} 
\and  Swati Ravi \inst{\ref{in:MIT}}\orcidlink{0000-0002-2381-4184} 
\and Herman~L.~Marshall \inst{\ref{in:MIT}}\orcidlink{0000-0002-6492-1293}  
\and Alessandro~Di~Marco \inst{\ref{in:INAF-IAPS}}\orcidlink{0000-0003-0331-3259} 
\and Fabio~La~Monaca \inst{\ref{in:INAF-IAPS},\ref{in:UniRoma2},\ref{in:LaSapienza}}\orcidlink{0000-0001-8916-4156}  
\and Christian~Malacaria \inst{\ref{in:ISSI}}\orcidlink{0000-0002-0380-0041} 
\and Guglielmo~Mastroserio \inst{\ref{in:Milano}}\orcidlink{0000-0003-4216-7936} 
\and Vladislav~Loktev \inst{\ref{in:UTU}}\orcidlink{0000-0001-6894-871X}  
\and Andrea~Possenti \inst{\ref{in:INAF-OAC}}\orcidlink{0000-0001-5902-3731}   
\and Valery~F.~Suleimanov\inst{\ref{in:Tub}}\orcidlink{0000-0003-3733-7267}  
\and Roberto~Taverna \inst{\ref{in:UniPD}}\orcidlink{0000-0002-1768-618X} 
\and Iv\'an~Agudo \inst{\ref{in:CSIC-IAA}}\orcidlink{0000-0002-3777-6182}  
\and Lucio~A.~Antonelli \inst{\ref{in:INAF-OAR},\ref{in:ASI-SSDC}}\orcidlink{0000-0002-5037-9034}   
\and Matteo~Bachetti \inst{\ref{in:INAF-OAC}}\orcidlink{0000-0002-4576-9337}   
\and Luca~Baldini  \inst{\ref{in:INFN-PI},	\ref{in:UniPI}}\orcidlink{0000-0002-9785-7726}   
\and Wayne~H.~Baumgartner  \inst{\ref{in:NASA-MSFC}}\orcidlink{0000-0002-5106-0463}  
\and Ronaldo~Bellazzini  \inst{\ref{in:INFN-PI}}\orcidlink{0000-0002-2469-7063}   
\and Stefano~Bianchi \inst{\ref{in:UniRoma3}}\orcidlink{0000-0002-4622-4240}  
\and Stephen~D.~Bongiorno \inst{\ref{in:NASA-MSFC}}\orcidlink{0000-0002-0901-2097}   
\and Raffaella~Bonino  \inst{\ref{in:INFN-TO},\ref{in:UniTO}}\orcidlink{0000-0002-4264-1215}  
\and Alessandro~Brez  \inst{\ref{in:INFN-PI}}\orcidlink{0000-0002-9460-1821}   
\and Niccol\`{o}~Bucciantini 
\inst{\ref{in:INAF-Arcetri},\ref{in:UniFI},\ref{in:INFN-FI}}\orcidlink{0000-0002-8848-1392}  
\and Fiamma~Capitanio \inst{\ref{in:INAF-IAPS}}\orcidlink{0000-0002-6384-3027} 
\and Simone~Castellano \inst{\ref{in:INFN-PI}}\orcidlink{0000-0003-1111-4292}   
\and Elisabetta~Cavazzuti \inst{\ref{in:ASI}}\orcidlink{0000-0001-7150-9638}   
\and Chien-Ting~Chen \inst{\ref{in:USRA-MSFC}}\orcidlink{0000-0002-4945-5079}  
\and Stefano~Ciprini \inst{\ref{in:INFN-Roma2},\ref{in:ASI-SSDC}}\orcidlink{0000-0002-0712-2479}  
\and Enrico~Costa \inst{\ref{in:INAF-IAPS}}\orcidlink{0000-0003-4925-8523}   
\and Alessandra~De~Rosa \inst{\ref{in:INAF-IAPS}}\orcidlink{0000-0001-5668-6863}  
\and Ettore~Del~Monte \inst{\ref{in:INAF-IAPS}}\orcidlink{0000-0002-3013-6334}   
\and Laura~Di~Gesu \inst{\ref{in:ASI}}\orcidlink{0000-0002-5614-5028}   
\and Niccol\`{o}~Di~Lalla \inst{\ref{in:Stanford}}\orcidlink{0000-0002-7574-1298}  
\and Immacolata~Donnarumma \inst{\ref{in:ASI}}\orcidlink{0000-0002-4700-4549}  
\and Michal~Dov\v{c}iak \inst{\ref{in:CAS-ASU}}\orcidlink{0000-0003-0079-1239}
\and Steven~R.~Ehlert \inst{\ref{in:NASA-MSFC}}\orcidlink{0000-0003-4420-2838}    
\and Teruaki~Enoto \inst{\ref{in:RIKEN}}\orcidlink{0000-0003-1244-3100}  
\and Yuri~Evangelista \inst{\ref{in:INAF-IAPS}}\orcidlink{0000-0001-6096-6710}  
\and Sergio~Fabiani \inst{\ref{in:INAF-IAPS}}\orcidlink{0000-0003-1533-0283}  
\and Riccardo~Ferrazzoli \inst{\ref{in:INAF-IAPS}}\orcidlink{0000-0003-1074-8605}   
\and Javier~A.~Garcia \inst{\ref{in:GSFC}}\orcidlink{0000-0003-3828-2448}
\and Shuichi~Gunji \inst{\ref{in:Yamagata}}\orcidlink{0000-0002-5881-2445}   
\and Kiyoshi~Hayashida \inst{\ref{in:Osaka}}\thanks{Deceased.}  
\and Jeremy~Heyl \inst{\ref{in:UBC}}\orcidlink{000-0001-9739-367X]} 
\and Wataru~Iwakiri \inst{\ref{in:Chiba}}\orcidlink{0000-0002-0207-9010}   
\and Svetlana~G.~Jorstad \inst{\ref{in:BU},\ref{in:SPBU}}\orcidlink{0000-0001-9522-5453}   
\and Philip~Kaaret \inst{\ref{in:NASA-MSFC}}\orcidlink{0000-0002-3638-0637} 
\and Vladimir~Karas \inst{\ref{in:CAS-ASU}}\orcidlink{0000-0002-5760-0459}  
\and Fabian~Kislat \inst{\ref{in:UNH}}\orcidlink{0000-0001-7477-0380}   
\and Takao~Kitaguchi  \inst{\ref{in:RIKEN}} 
\and Jeffery~J.~Kolodziejczak \inst{\ref{in:NASA-MSFC}}\orcidlink{0000-0002-0110-6136}   
\and Henric~Krawczynski \inst{\ref{in:StLW}}\orcidlink{0000-0002-1084-6507} 
\and Luca~Latronico  \inst{\ref{in:INFN-TO}}\orcidlink{0000-0002-0984-1856}   
\and Ioannis~Liodakis \inst{\ref{in:NASA-MSFC},\ref{in:AstroCrete}}\orcidlink{0000-0001-9200-4006}
\and Simone~Maldera \inst{\ref{in:INFN-TO}}\orcidlink{0000-0002-0698-4421}    
\and Alberto~Manfreda \inst{\ref{INFN-NA}}\orcidlink{0000-0002-0998-4953}  
\and Fr\'{e}d\'{e}ric~Marin \inst{\ref{in:Strasbourg}}\orcidlink{0000-0003-4952-0835}   
\and Andrea~Marinucci \inst{\ref{in:ASI}}\orcidlink{0000-0002-2055-4946}   
\and Alan~P.~Marscher \inst{\ref{in:BU}}\orcidlink{0000-0001-7396-3332}   
\and Francesco~Massaro \inst{\ref{in:INFN-TO},\ref{in:UniTO}}\orcidlink{0000-0002-1704-9850}   
\and Giorgio~Matt  \inst{\ref{in:UniRoma3}}\orcidlink{0000-0002-2152-0916}  
\and Ikuyuki~Mitsuishi \inst{\ref{in:Nagoya}}
\and Tsunefumi~Mizuno \inst{\ref{in:Hiroshima}}\orcidlink{0000-0001-7263-0296}   
\and Fabio~Muleri \inst{\ref{in:INAF-IAPS}}\orcidlink{0000-0003-3331-3794}
\and Michela~Negro \inst{\ref{in:LSU}}\orcidlink{0000-0002-6548-5622} 
\and Chi-Yung~Ng \inst{\ref{in:HKU}}\orcidlink{000-0002-5847-2612}  
\and Stephen~L.~O'Dell \inst{\ref{in:NASA-MSFC}}\orcidlink{0000-0002-1868-8056}    
\and Nicola~Omodei \inst{\ref{in:Stanford}}\orcidlink{0000-0002-5448-7577}  
\and Chiara~Oppedisano \inst{\ref{in:INFN-TO}}\orcidlink{0000-0001-6194-4601}    
\and Alessandro~Papitto \inst{\ref{in:INAF-OAR}}\orcidlink{0000-0001-6289-7413}  
\and George~G.~Pavlov \inst{\ref{in:PSU}}\orcidlink{0000-0002-7481-5259}   
\and Abel~L.~Peirson \inst{\ref{in:Stanford}}\orcidlink{0000-0001-6292-1911}  
\and Matteo~Perri \inst{\ref{in:ASI-SSDC},\ref{in:INAF-OAR}}\orcidlink{0000-0003-3613-4409}  
\and Melissa~Pesce-Rollins \inst{\ref{in:INFN-PI}}\orcidlink{0000-0003-1790-8018}   
\and Pierre-Olivier~Petrucci \inst{\ref{in:Grenoble}}\orcidlink{0000-0001-6061-3480}
\and Maura~Pilia \inst{\ref{in:INAF-OAC}}\orcidlink{0000-0001-7397-8091}   
\and Simonetta~Puccetti \inst{\ref{in:ASI-SSDC}}\orcidlink{0000-0002-2734-7835}  
\and Brian~D.~Ramsey \inst{\ref{in:NASA-MSFC}}\orcidlink{0000-0003-1548-1524}    
\and John~Rankin \inst{\ref{in:INAF-IAPS}}\orcidlink{0000-0002-9774-0560}   
\and Ajay~Ratheesh \inst{\ref{in:INAF-IAPS}}\orcidlink{0000-0003-0411-4243} 
\and Oliver~J.~Roberts \inst{\ref{in:USRA-MSFC}}\orcidlink{0000-0002-7150-9061}  
\and Roger~W.~Romani \inst{\ref{in:Stanford}}\orcidlink{0000-0001-6711-3286}  
\and Carmelo~Sgr\`o \inst{\ref{in:INFN-PI}}\orcidlink{0000-0001-5676-6214}    
\and Patrick~Slane \inst{\ref{in:CfA}}\orcidlink{0000-0002-6986-6756}    
\and Paolo~Soffitta \inst{\ref{in:INAF-IAPS}}\orcidlink{0000-0002-7781-4104}   
\and Gloria~Spandre \inst{\ref{in:INFN-PI}}\orcidlink{0000-0003-0802-3453}   
\and Douglas~A.~Swartz \inst{\ref{in:USRA-MSFC}}\orcidlink{0000-0002-2954-4461}  
\and Toru~Tamagawa \inst{\ref{in:RIKEN}}\orcidlink{0000-0002-8801-6263}  
\and Fabrizio~Tavecchio \inst{\ref{in:INAF-OAB}}\orcidlink{0000-0003-0256-0995}  
\and Yuzuru~Tawara \inst{\ref{in:Nagoya}} 
\and Allyn~F.~Tennant \inst{\ref{in:NASA-MSFC}}\orcidlink{0000-0002-9443-6774}    
\and Nicholas~E.~Thomas \inst{\ref{in:NASA-MSFC}}\orcidlink{0000-0003-0411-4606}    
\and Francesco~Tombesi  \inst{\ref{in:UniRoma2},\ref{in:INFN-Roma2}}\orcidlink{0000-0002-6562-8654}  
\and Alessio~Trois \inst{\ref{in:INAF-OAC}}\orcidlink{0000-0002-3180-6002}  
\and Roberto~Turolla \inst{\ref{in:UniPD},\ref{in:MSSL}}\orcidlink{0000-0003-3977-8760}  
\and Jacco~Vink \inst{\ref{in:Amsterdam}}\orcidlink{0000-0002-4708-4219}  
\and Martin~C.~Weisskopf \inst{\ref{in:NASA-MSFC}}\orcidlink{0000-0002-5270-4240}   
\and Kinwah~Wu \inst{\ref{in:MSSL}}\orcidlink{0000-0002-7568-8765}  
\and Fei~Xie \inst{\ref{in:GSU},\ref{in:INAF-IAPS}}\orcidlink{0000-0002-0105-5826}  
\and Silvia~Zane  \inst{\ref{in:MSSL}}\orcidlink{0000-0001-5326-880X}  
}

\institute{Department of Physics and Astronomy, FI-20014 University of Turku,  Finland \label{in:UTU} \\ \email{sofia.v.forsblom@utu.fi}
\and 
Institut f\"ur Astronomie und Astrophysik, Universit\"at T\"ubingen, Sand 1, D-72076 T\"ubingen, Germany \label{in:Tub} 
\and 
Astrophysics, Department of Physics, University of Oxford, Denys Wilkinson Building, Keble Road, Oxford OX1 3RH, UK \label{in:Oxford} 
\and 
MIT Kavli Institute for Astrophysics and Space Research, Massachusetts Institute of Technology, 77 Massachusetts Avenue, Cambridge, MA 02139, USA \label{in:MIT}
\and  INAF Istituto di Astrofisica e Planetologia Spaziali, Via del Fosso del Cavaliere 100, 00133 Roma, Italy \label{in:INAF-IAPS} 
\and
Dipartimento di Fisica, Universit\`{a} degli Studi di Roma ``Tor Vergata'', Via della Ricerca Scientifica 1, 00133 Roma, Italy \label{in:UniRoma2}
\and 
Dipartimento di Fisica, Universit\`{a} degli Studi di Roma ``La Sapienza'', Piazzale Aldo Moro 5, 00185 Roma, Italy \label{in:LaSapienza} 
\and International Space Science Institute, Hallerstrasse 6, 3012 Bern, Switzerland \label{in:ISSI} 
\and Dipartimento di Fisica, Universit\`{a} degli Studi di Milano, Via Celoria 16, I-20133 Milano, Italy \label{in:Milano} 
\and INAF Osservatorio Astronomico di Cagliari, Via della Scienza 5, 09047 Selargius (CA), Italy  \label{in:INAF-OAC} 
\and 
Dipartimento di Fisica e Astronomia, Universit\`{a} degli Studi di Padova, Via Marzolo 8, 35131 Padova, Italy \label{in:UniPD}
\and 
Instituto de Astrof\'{i}sica de Andaluc\'{i}a -- CSIC, Glorieta de la Astronom\'{i}a s/n, 18008 Granada, Spain \label{in:CSIC-IAA}
\and 
INAF Osservatorio Astronomico di Roma, Via Frascati 33, 00040 Monte Porzio Catone (RM), Italy \label{in:INAF-OAR} 	
\and 
Space Science Data Center, Agenzia Spaziale Italiana, Via del Politecnico snc, 00133 Roma, Italy \label{in:ASI-SSDC}
\and 
Istituto Nazionale di Fisica Nucleare, Sezione di Pisa, Largo B. Pontecorvo 3, 56127 Pisa, Italy \label{in:INFN-PI}
\and  
Dipartimento di Fisica, Universit\`{a} di Pisa, Largo B. Pontecorvo 3, 56127 Pisa, Italy \label{in:UniPI} 
\and NASA Marshall Space Flight Center, Huntsville, AL 35812, USA \label{in:NASA-MSFC}
\and 
Dipartimento di Matematica e Fisica, Universit\`a degli Studi Roma Tre, via della Vasca Navale 84, 00146 Roma, Italy  \label{in:UniRoma3}
\and  
Istituto Nazionale di Fisica Nucleare, Sezione di Torino, Via Pietro Giuria 1, 10125 Torino, Italy  \label{in:INFN-TO}	
\and  
Dipartimento di Fisica, Universit\`{a} degli Studi di Torino, Via Pietro Giuria 1, 10125 Torino, Italy \label{in:UniTO} 
\and   
INAF Osservatorio Astrofisico di Arcetri, Largo Enrico Fermi 5, 50125 Firenze, Italy 
\label{in:INAF-Arcetri} 
\and  
Dipartimento di Fisica e Astronomia, Universit\`{a} degli Studi di Firenze, Via Sansone 1, 50019 Sesto Fiorentino (FI), Italy \label{in:UniFI} 
\and   
Istituto Nazionale di Fisica Nucleare, Sezione di Firenze, Via Sansone 1, 50019 Sesto Fiorentino (FI), Italy \label{in:INFN-FI}
\and 
Agenzia Spaziale Italiana, Via del Politecnico snc, 00133 Roma, Italy \label{in:ASI}
\and 
Science and Technology Institute, Universities Space Research Association, Huntsville, AL 35805, USA \label{in:USRA-MSFC}
\and 
Istituto Nazionale di Fisica Nucleare, Sezione di Roma ``Tor Vergata'', Via della Ricerca Scientifica 1, 00133 Roma, Italy 
 \label{in:INFN-Roma2}
\and 
Department of Physics and Kavli Institute for Particle Astrophysics and Cosmology, Stanford University, Stanford, California 94305, USA  \label{in:Stanford}
\and 
Astronomical Institute of the Czech Academy of Sciences, Boční II 1401/1, 14100 Praha 4, Czech Republic \label{in:CAS-ASU}
\and 
RIKEN Cluster for Pioneering Research, 2-1 Hirosawa, Wako, Saitama 351-0198, Japan \label{in:RIKEN}
\and 
X-ray Astrophysics Laboratory, NASA Goddard Space Flight Center, Greenbelt, MD 20771, USA \label{in:GSFC}
\and 
Yamagata University,1-4-12 Kojirakawa-machi, Yamagata-shi 990-8560, Japan \label{in:Yamagata}
\and 
Osaka University, 1-1 Yamadaoka, Suita, Osaka 565-0871, Japan \label{in:Osaka}
\and 
University of British Columbia, Vancouver, BC V6T 1Z4, Canada \label{in:UBC}
\and 
International Center for Hadron Astrophysics, Chiba University, Chiba 263-8522, Japan \label{in:Chiba}
\and
Institute for Astrophysical Research, Boston University, 725 Commonwealth Avenue, Boston, MA 02215, USA \label{in:BU}  
\and 
Department of Astrophysics, St. Petersburg State University, Universitetsky pr. 28, Petrodvoretz, 198504 St. Petersburg, Russia \label{in:SPBU}
\and 
Department of Physics and Astronomy and Space Science Center, University of New Hampshire, Durham, NH 03824, USA \label{in:UNH} 
\and 
Physics Department and McDonnell Center for the Space Sciences, Washington University in St. Louis, St. Louis, MO 63130, USA \label{in:StLW}
\and 
Institute of Astrophysics, Foundation for Research and Technology-Hellas, GR-70013 Heraklion, Greece \label{in:AstroCrete} 
\and 
Istituto Nazionale di Fisica Nucleare, Sezione di Napoli, Strada Comunale Cinthia, 80126 Napoli, Italy \label{INFN-NA}
\and 
Universit\'{e} de Strasbourg, CNRS, Observatoire Astronomique de Strasbourg, UMR 7550, 67000 Strasbourg, France \label{in:Strasbourg}
\and 
Graduate School of Science, Division of Particle and Astrophysical Science, Nagoya University, Furo-cho, Chikusa-ku, Nagoya, Aichi 464-8602, Japan \label{in:Nagoya}
\and 
Hiroshima Astrophysical Science Center, Hiroshima University, 1-3-1 Kagamiyama, Higashi-Hiroshima, Hiroshima 739-8526, Japan \label{in:Hiroshima}
\and 
Department of Physics and Astronomy, Louisiana State University, Baton Rouge, LA 70803, USA \label{in:LSU} 
\and 
Department of Physics, University of Hong Kong, Pokfulam, Hong Kong \label{in:HKU}
\and 
Department of Astronomy and Astrophysics, Pennsylvania State University, University Park, PA 16801, USA \label{in:PSU}
\and 
Universit\'{e} Grenoble Alpes, CNRS, IPAG, 38000 Grenoble, France \label{in:Grenoble}
\and 
Center for Astrophysics, Harvard \& Smithsonian, 60 Garden St, Cambridge, MA 02138, USA \label{in:CfA} 
\and 
INAF Osservatorio Astronomico di Brera, via E. Bianchi 46, 23807 Merate (LC), Italy \label{in:INAF-OAB}
\and 
Mullard Space Science Laboratory, University College London, Holmbury St Mary, Dorking, Surrey RH5 6NT, UK \label{in:MSSL}
\and 
Anton Pannekoek Institute for Astronomy \& GRAPPA, University of Amsterdam, Science Park 904, 1098 XH Amsterdam, The Netherlands  \label{in:Amsterdam} 
\and 
Guangxi Key Laboratory for Relativistic Astrophysics, School of Physical Science and Technology, Guangxi University, Nanning 530004, China \label{in:GSU}
} 
  
\titlerunning{X-ray polarimetry of the X-ray pulsar \smc}
\authorrunning{S.~V.~Forsblom et al.}

\date{2024}

\abstract{Recent observations of X-ray pulsars (XRPs) performed by the {Imaging X-ray Polarimetry Explorer} (\ixpe) have made it possible to investigate the intricate details of these objects in a new way, thanks to the added value of X-ray polarimetry.
Here we present the results of the \ixpe observations of \smc, a member of the small group of XRPs displaying super-orbital variability.
\smc was observed by \ixpe three separate times during the high state of its super-orbital period.
The observed luminosity in the 2--8~keV energy band of $L\sim2\times10^{38}$\,erg\,s$^{-1}$ makes \smc the brightest XRP ever observed by \ixpe.  
We detect significant polarization in all three observations, with values of the phase-averaged polarization degree (PD) and polarization angle (PA) of $3.2\pm 0.8$\% and $97\degr\pm 8\degr$ for Observation 1, $3.0\pm 0.9$\% and $90\degr\pm 8\degr$ for Observation 2, and $5.5\pm 1.1$\% and $80\degr\pm 6\degr$ for Observation 3, for the spectro-polarimetric analysis.
The observed PD shows an increase over time with decreasing luminosity, while the PA decreases in decrements of $\sim$10\degr.
The phase-resolved spectro-polarimetric analysis reveals significant detection of polarization in three out of seven phase bins, with the PD ranging between $\sim$2\% and $\sim$10\%, and a corresponding range in the PA from  $\sim$70\degr\ to $\sim$100\degr.
The pulse-phase resolved PD displays an apparent anti-correlation with the flux.
Using the rotating vector model, we obtain constraints on the pulsar’s geometrical properties for the individual observations. The position angle of the pulsar displays an evolution over time supporting the idea that we observe changes  related to different super-orbital phases. Scattering in the wind of the precessing accretion disk may be responsible for the behavior of the polarimetric properties observed during the high-state of \smc's super-orbital period.
}

\keywords{accretion, accretion disks -- magnetic fields -- polarization -- pulsars: individual: SMC~X-1 -- stars: neutron -- X-rays: binaries}

\maketitle
%
\section{Introduction} 
\label{sec:intro}

High-mass X-ray Binaries (HMXBs) involve a compact object accreting matter from a massive companion star.
For strongly magnetized neutron stars (NSs), the accretion flow is disrupted by the magnetosphere at a distance of the order of $10^{8}-10^{9}$\,cm. The matter follows then the magnetic field lines onto the NSs surface, where the gravitational potential energy of the accreted matter is released. Depending on the accretion rate, either hot spots or extended accretion columns may form at the magnetic poles. 
The misalignment between the magnetic poles and spin axis leads to pulsed X-ray emission, and the appearance of an X-ray pulsar (XRP; for a recent review, see, e.g., \citealt{MushtukovTsygankov2024}).
The observed properties and phase-dependence of the pulsed X-ray emission are defined by the emission region geometry, the pulsar's orientation with respect to the observer, and details of radiative transfer in the strong magnetic field. There are, however, no robust theoretical models capable of describing those comprehensively. 
One of the main issues is the uncertainty in the basic geometry of XRPs. 
Polarimetric observations can be used to determine the pulsar geometry (e.g., the inclination of the pulsar spin to the line of sight and the magnetic obliquity) and possibly to distinguish the emission region geometry.

The immense magnetic field of the NS and its impact on Compton scattering cross-sections are the main reasons for the polarized X-ray emission from XRPs. The scattering of photons in highly magnetized plasma is expected to result in a large degree of polarization (up to 80\%) of the emerging X-ray emission for favorable orientations.
It has been shown \citep{Meszaros88} that the linear X-ray polarization is strongly dependent on the geometry of the emission region and variable with energy and pulse phase, and phase-resolved polarimetry can be used to constrain the viewing geometry and discern different radiation models.
\ixpe\ has made it possible to detect the linear X-ray polarization for several tens of astrophysical X-ray sources including XRPs: \mbox{Her~X-1} \citep{2022NatAs...6.1433D,Heyl24,2024MNRAS.tmp.1187Z},  \mbox{Cen~X-3} \citep{2022ApJ...941L..14T}, X~Persei \citep{2023MNRAS.524.2004M}, \mbox{4U~1626$-$67} \citep{2022ApJ...940...70M}, \mbox{Vela~X-1} \citep{2023ApJ...947L..20F}, \mbox{GRO~J1008$-$57} \citep{2023A&A...675A..48T}, EXO~2030+375 \citep{2023A&A...675A..29M}, \mbox{LS~V~+44~17} \citep{2023A&A...677A..57D}, \mbox{GX~301$-$2} \citep{Suleimanov2023}, and \mbox{Swift~J0243.6+6124} \citep{Poutanen24}. 
The geometry of the emission region depends on the accretion luminosity, and two main emission region geometries can be distinguished based on the local mass accretion rate: a hot spot and an accretion column. The critical luminosity separating these two accretion regimes is a function of the physical and geometrical parameters of the system. 
The XRPs observed by \ixpe so far have all been in the sub-critical regime, making observations of super-critical XRPs important objects to supplement the information of the polarization properties of XRPs in general. Polarimetric observations of the bright XRP \smc can help provide information on the super-critical regime and shed light on the effects of  accretion columns on the observed polarization.

\smc, a well-studied HMXB in the Small Magellanic Cloud, was initially detected by \citet{1971-Price} and later confirmed as a discrete source by \citet{1971-Leong}, who observed significant variability in the intensity and spectrum. \citet{1972-Schreier} verified its binary nature, discovering periodic occultations with a $\sim$3.9~d orbital period and measuring the binary orbit inclination of $\sim$70\degr. \smc\ also exhibits pulsations with the period of about 0.7~s, with varying pulse characteristics over time \citep{1976-Lucke}.
\smc, one of a few super-giant X-ray binaries known for Roche lobe overflow accretion, consistently emits near or above its Eddington luminosity, which is around $1.3\times10^{38}$ \lumcgs\ for an estimated mass of 1.1$M_{\odot}$ for the NS in the system \citep{2007-vanderMeer}. 
Its apparent luminosity varies between approximately $10^{37}$ \lumcgs\ in the low state to over $5\times10^{38}$ \lumcgs\ in the high state, more than three times its Eddington luminosity \citep{1981-Bonnet-Bidaud}. 
In addition to this persistent emission, \smc\ also displays type II X-ray bursts lasting tens of seconds \citep{1991-Angelini,2018-Rai}. 
The near- to super-Eddington luminosity places \smc\ between less luminous Be/X-ray binaries \citep{2011-Reig} and brighter ultra-luminous XRPs \citep{2017ARA&A..55..303K}. 

\smc\ is also part of an important, but small, group of XRPs that display super-orbital variability, believed to be caused by a warped and precessing accretion disk.
The precession of the disk causes periodic, partial obscuration of the central source, giving rise to the modulation of the X-ray flux at the precession period.
The super-orbital period of \smc\ is not steady, as noted by \citet{1998-Wojdowski}.
An instability in the warped accretion disk, changing its geometry as it cycles between stable modes, is believed to cause so-called ``excursions'', during which the super-orbital period decreases from its average value of 55~d to $\sim$40~d \citep{2003-Clarkson,2019-Dage,2013-Hu}.
The precession of the disk in \smc\ gives rise to three distinct super-orbital states: the high-state characterized by maximum source flux, the low-state of minimum flux (caused by the occultation of the NS by the disk), and the intermediate state marking the transition between the high and low state.
\smc represents a special case where the geometry of the large-scale accretion flow is interconnected with the emission region of the XRP in the very vicinity of the NS.
X-ray polarimetry allows probing the emission geometry, and hence, is the perfect tool to study the coupling between the accretion disk and the NS.

\begin{figure}
\centering
\includegraphics[width=1.0\columnwidth]{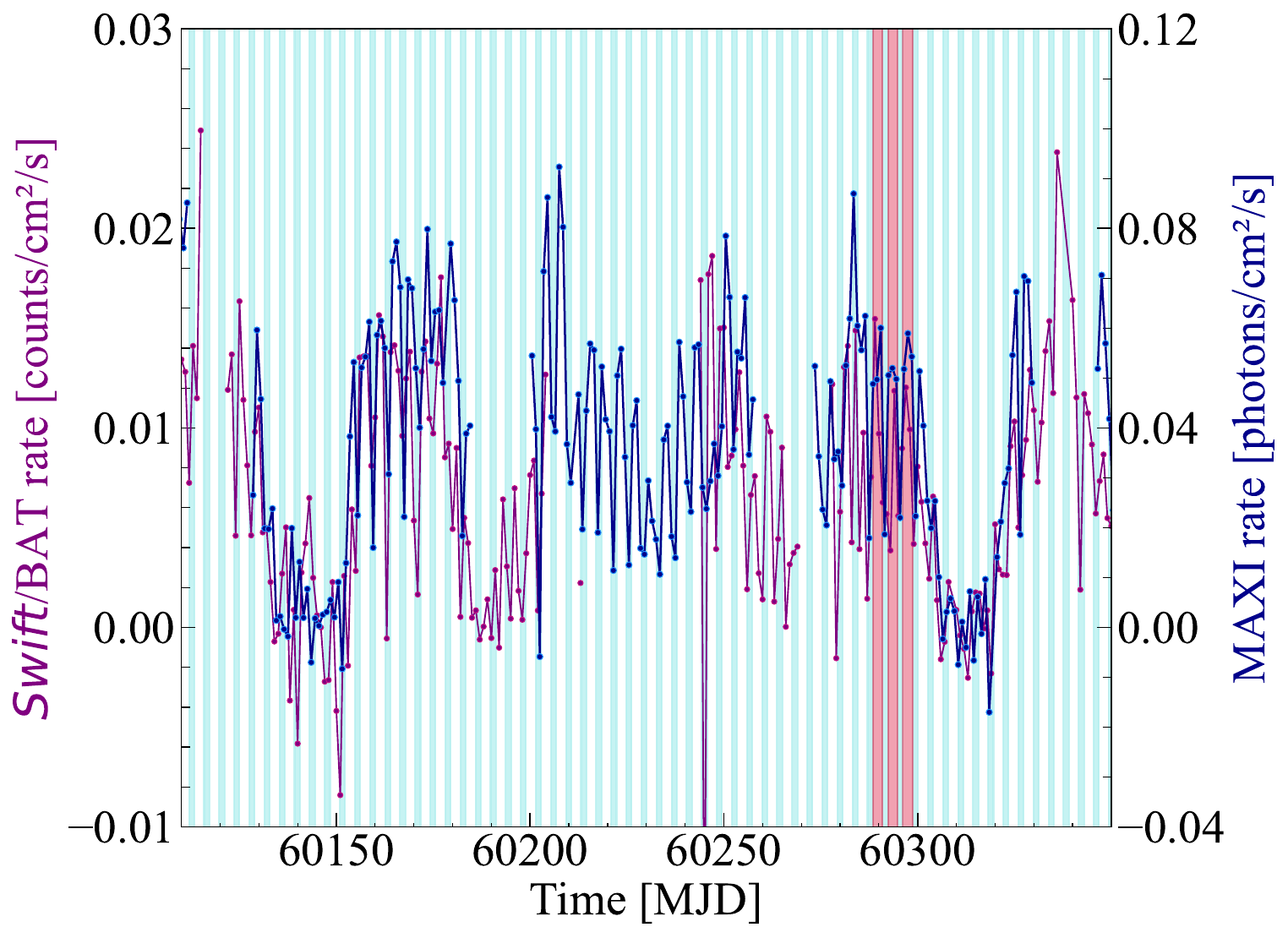}
\caption{\textit{Swift}/BAT (15--50 keV) and MAXI (4--10 keV) one-day averaged light curves of \smc in purple and blue, respectively. 
Vertical light blue lines show the eclipses and
vertical pink lines display the times of the observations with \ixpe.
Error bars have been removed for visual clarity.}
\label{fig:swift-lc}
\end{figure}
\begin{figure}
\centering
\includegraphics[width=1.0\columnwidth]{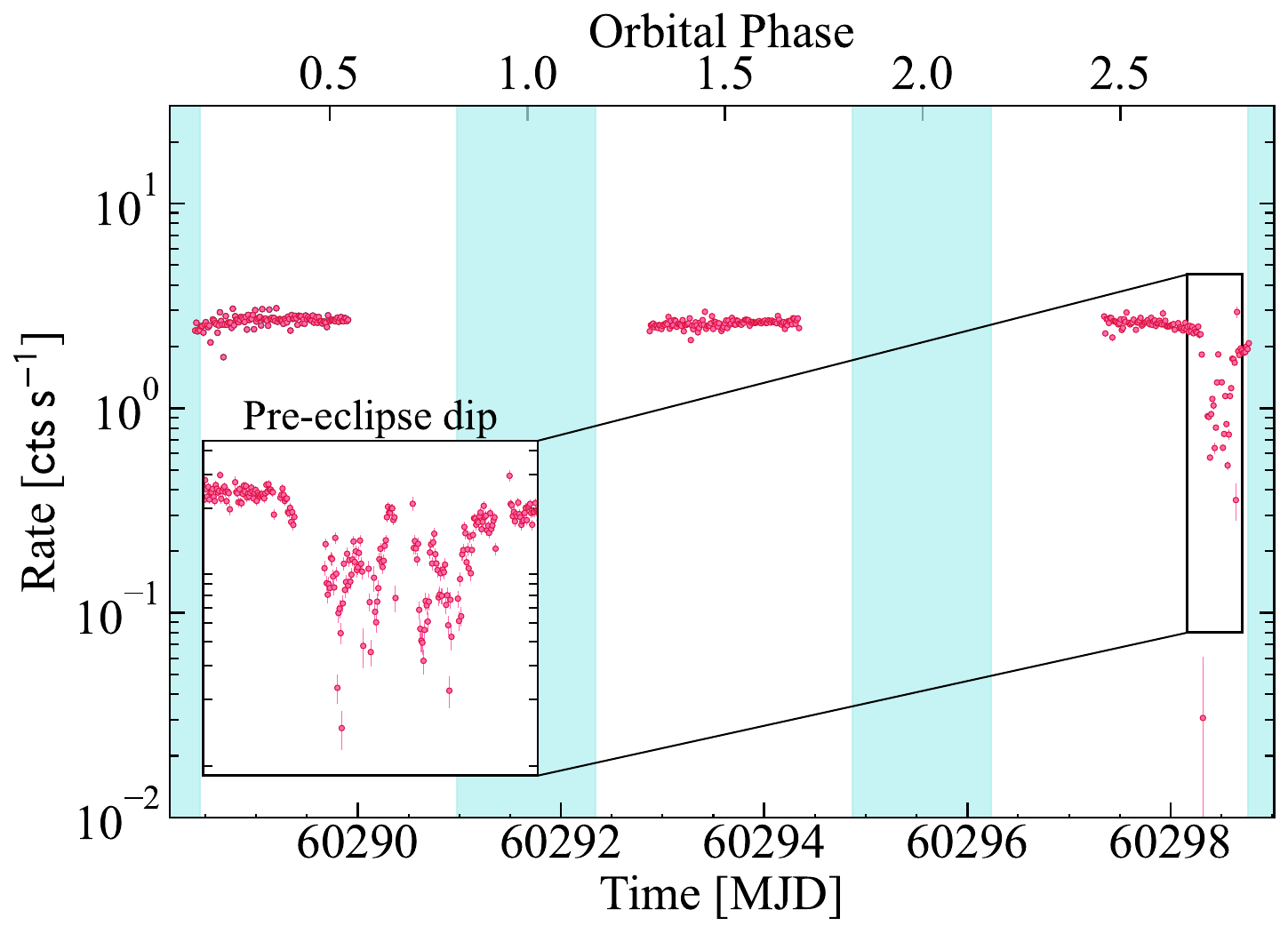}
\caption{Light curve of \smc observed with \ixpe in the 2--8 keV energy band. Times of eclipses are shown in the blue shaded regions. The inset displays a higher time-resolution light curve of the pre-eclipse dip detected during the third observation.}
\label{fig:ixpe-lc}
\end{figure}

\begin{figure*}
\centering
\includegraphics[width=0.66\columnwidth]{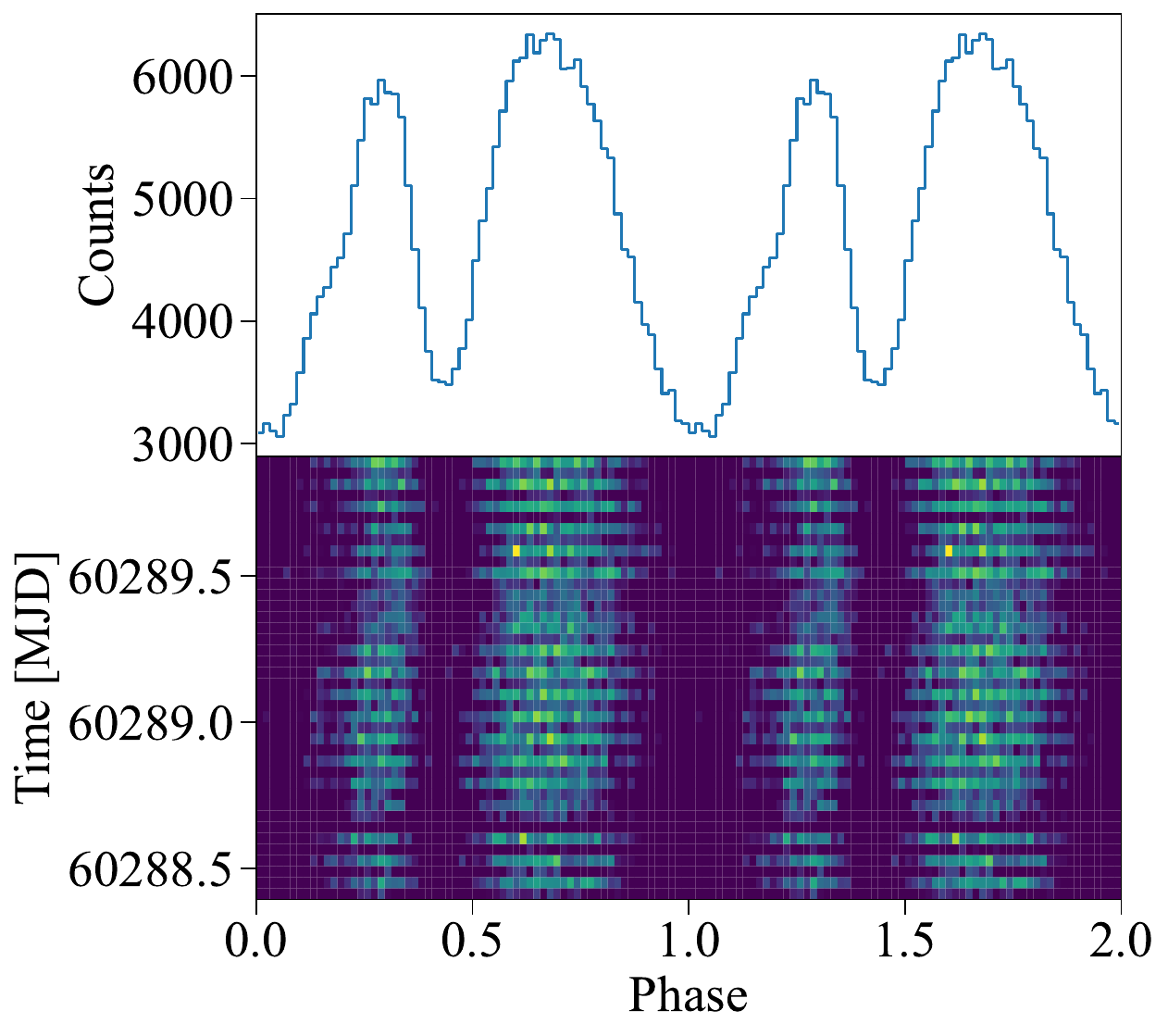}
\includegraphics[width=0.66\columnwidth]{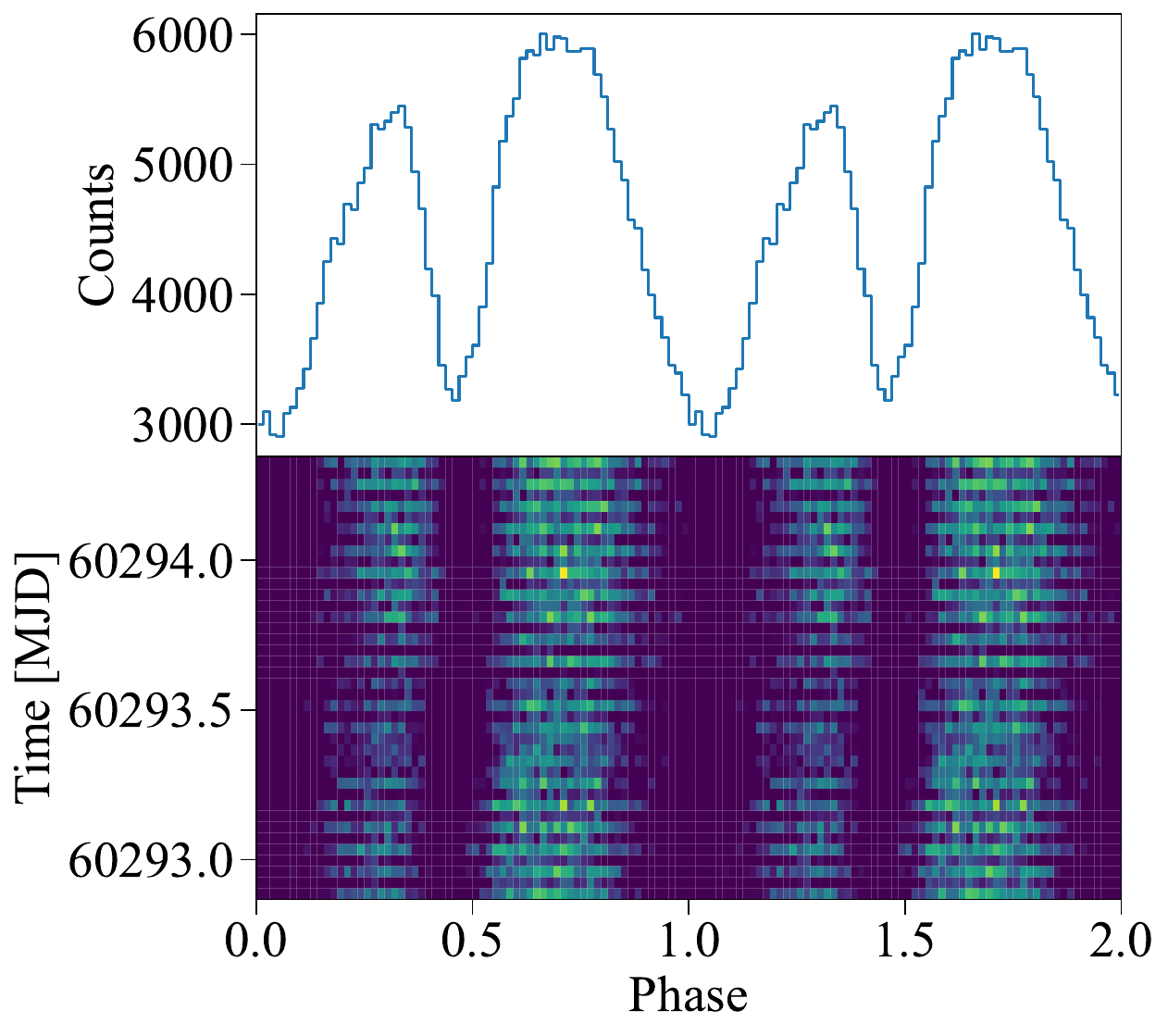}
\includegraphics[width=0.66\columnwidth]{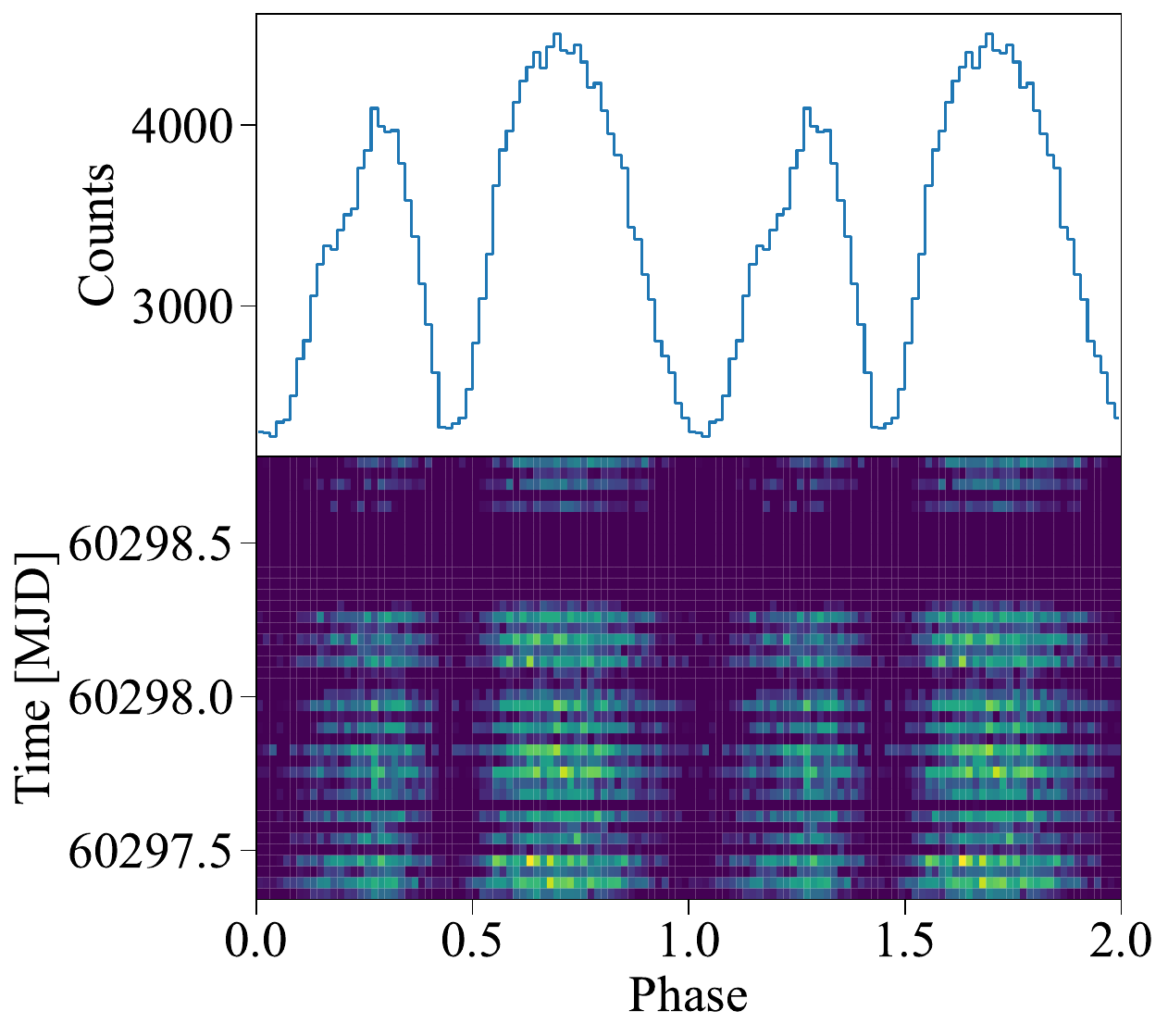}
\caption{Phase-aligned pulse profiles and phaseograms for \smc as seen by \ixpe in the 2--8 keV energy band for the Obs.~1 (\textit{left}), Obs.~2 (\textit{center}), and Obs.~3 (\textit{right}). }
\label{fig:ixpe-pulseprofile}
\end{figure*}

\begin{table*} 
\centering
\caption{Timing parameters used for Obs. 1, 2, and 3 of \smc.
} 
\begin{tabular}{llccc}
\hline
\hline
 Parameter & Unit & Obs. 1 & Obs. 2 & Obs. 3\\
\hline
        Epoch & MET\tablefootmark{a} & 218971574.938 & 219358107.288 & 219744614.375  \\
        Spin period & ms & 696.6750$\pm0.0003$ & 696.6712$\pm0.0003$ & 696.6659$\pm0.0004$  \\
        Spin period derivative & $10^{-11}\mathrm{s\;s^{-1}}$ & $-1.3\pm0.4$ & $-2.9\pm0.4$ & $-3.0\pm1.0$  \\
\hline  
\end{tabular}
\tablefoot{The timing solutions refer to the individual observations.
\tablefoottext{a}{\ixpe MET (Mission Elapsed Time, in seconds) is equal to $T_{\rm MET}=(T_{\rm MJD}-T_{\rm ref})\times 86400$, where $T_{\rm ref}=57754.00080074074$. The reported epochs define the zero phases of the pulse profiles in Fig.~\ref{fig:ixpe-pulseprofile}. }}
\label{table:time-pars}
\end{table*}



\section{Observations and data reduction} 
\label{sec:data}

\ixpe is an observatory launched in December 2021 as a NASA/ASI mission \citep{Weisskopf2022}, with the goal of providing imaging polarimetry over the 2--8 keV energy range.
\ixpe is made up of three grazing incidence telescopes, each consisting of a mirror module assembly (MMA), which focuses X-rays onto a focal-plane polarization-sensitive gas pixel detector unit \citep[DU;][]{2021AJ....162..208S,2021APh...13302628B}.
In addition to measuring the sky coordinates, time of arrival, and energy of each detected photon, it also measures the direction of the photo-electron which allows for polarimetric analysis.

\begin{table*}
\centering
\caption{Measurements of the normalized Stokes parameters $q$ and $u$, PD, and PA for the phase-averaged data of \smc for different intervals using the \texttt{pcube} algorithm. 
}
\begin{tabular}{lcccc}
    \hline\hline
    Interval  & $q$ & $u$ & PD & PA \\ 
          & (\%) & (\%) & (\%) & (deg) \\ 
    \hline
    Obs. 1 & $-2.4\pm1.2$ & $-1.0\pm1.2$ & $2.6\pm1.2$ & $101\pm13$ \\
    Obs. 2 & $-3.6\pm1.2$ & $0.2\pm1.2$ & $3.6\pm1.2$ & $88\pm10$ \\
    Obs. 3 (incl. dip) & $-5.1\pm1.4$ & $1.8\pm1.4$ & $5.5\pm1.4$ & $80\pm7$ \\
    Obs. 3 (excl. dip) & $-6.3\pm1.5$ & $2.1\pm1.5$ & $6.6\pm1.5$ & $81\pm6$ \\
            \hline
    Obs. 1--3   (incl. dip) & $-3.6\pm0.7$ & $0.2\pm0.7$ & $3.6\pm0.7$ & $88\pm6$ \\
    Obs. 1--3  (excl. dip) & $-3.8\pm0.7$ & $0.2\pm0.7$ & $3.8\pm0.7$ & $89\pm6$ \\
            \hline    
    Pre-eclipse dip & $0.3\pm3.8$ & $0.3\pm3.8$ & $0.4\pm3.8$ & unconstrained\tablefootmark{a} \\
    \hline
    \end{tabular}
    \tablefoot{\tablefoottext{a}{Formal value for the PA is $20\degr\pm250$\degr.}}
    \label{table:avebins}
\end{table*}

\ixpe observed \smc in December 2023, during the high-state of the super-orbital period (during three consecutive orbital periods, see Fig.~\ref{fig:swift-lc}), for a total exposure of $\sim$320 ks.
The light curve in the 2--8 keV energy range from the \ixpe observations of \smc is shown in Fig.~\ref{fig:ixpe-lc}.
Three observations (hereinafter referred to as Obs.~1, 2, and 3) occurred between December 10--11, 14--16, and 19--20, with total effective exposures of 111, 110, and 97~ks, respectively.
Data have been processed with the {\sc ixpeobssim} package version 31.0.1 using the CalDB released on 2024 February 28.
The position offset correction and energy calibration were applied before the data analysis.
Source photons were extracted from a circular region centered on the source, with the radius $R_{\rm src}=80\arcsec$.
Due to the brightness of the source, background subtraction was not applied and the unweighted approach was used \citep{Di_Marco_2022,Di_Marco_2023}.

Event arrival times were corrected to the barycenter of the solar system using the \texttt{barycorr} tool from the {\sc ftools} package.
To account for the effects of binary orbital motion, we corrected observed event arrival times using the ephemerides by \citet{2010-Raichur-orb} and \citet{2019-Hu-orb} extrapolated to the \ixpe epoch. 
In particular, the latter is used to estimate the \ixpe epoch for mean longitude of 90\degr, which is estimated at MJD 60287.8042(2)  considering reported uncertainties for the zero epoch, the orbital period, and its derivative.
We found, however, that even after correction, the full set of \ixpe observations containing three orbital cycles could not be described with a single polynomial timing solution, i.e. some residuals were observed in pulse arrival times, hence the spin evolution of the source is more complex. 
Considering the presence of gaps in the data, describing that properly is challenging, so we opted to search for the spin period and its derivative using ${Z^2}$ statistics in each of the \ixpe segments separately, further refining the results using phase-connection. The values for spin period, its derivative, and the pulse epoch for each segment are given in Table~\ref{table:time-pars}. 
This approach allows us to obtain high-quality timing solutions with no phase drifts for each segment as illustrated in Fig.~\ref{fig:ixpe-pulseprofile}. 

Stokes $I$, $Q$, and $U$ spectra have been re-binned to have at least 30 counts per energy channel, with the same energy-binning applied to all energy spectra.
The energy spectra were fitted simultaneously using the {\sc xspec} package (version 12.14.0) \citep{Arn96} using $\chi^2$ statistics and the version 13 instrument response functions (ixpe:obssim20230702:v13).
The reported uncertainties are at the 68.3\% confidence level (1$\sigma$) unless stated otherwise.

\section{Results} 
\label{sec:res}

\subsection{Light curve and pulse profile}

\smc was observed three separate times during 2023 December 10--20, corresponding to the high-state of the super-orbital period as seen in the light curve (see Fig.~\ref{fig:swift-lc}) obtained by the \textit{Swift}/BAT\footnote{\url{https://swift.gsfc.nasa.gov/results/transients/}} \citep{Gehrels04} monitor and MAXI.\footnote{\url{http://maxi.riken.jp/}} 
During the third observation, a pre-eclipse dip can be seen in the \ixpe light curve (see Fig.~\ref{fig:ixpe-lc}), as previously observed for \smc during similar orbital phases \citep{2003-Moon,2007-Trowbridge,2013-Hu}.
The spin period and spin period derivative were measured for each observation separately and the determined values can be found in Table~\ref{table:time-pars}. 
The resulting pulse profiles for the observations of \smc in the 2--8 keV energy band are shown in Fig.~\ref{fig:ixpe-pulseprofile}, together with the phaseograms which display the evolution of the pulse profile over the time of each observation.


\begin{figure}
\centering
\includegraphics[width=0.84\columnwidth]{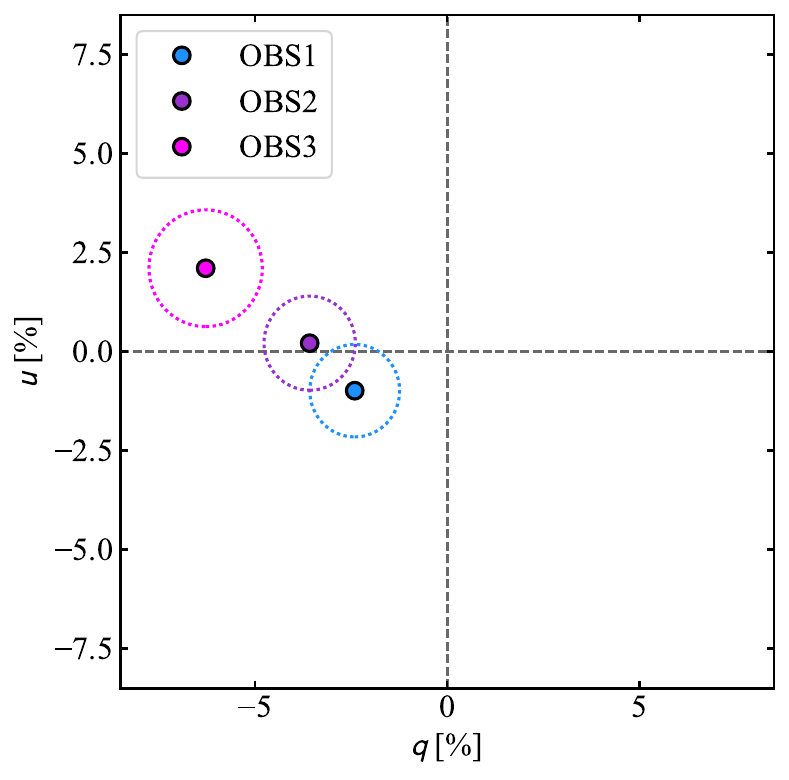}
\caption{
Phase-averaged normalized Stokes \textit{q} and \textit{u} of Observations 1, 2, and 3 (excluding dip) for each separate observation (combining the DUs) for the entire 2--8 keV energy band. The size of the circles correspond to the uncertainty at 68\% confidence level.}
\label{fig:timeres}
\end{figure}

\begin{table}
\centering
\caption{Measurements of the normalized Stokes parameters $q$ and $u$, PD, and PA in different energy bins using the \texttt{pcube} algorithm. 
}
\begin{tabular}{ccccc}
    \hline\hline
    Energy  & $q$ & $u$ & PD & PA \\ 
       (keV)   & (\%) & (\%) & (\%) & (deg) \\ 
    \hline
    2--3 & $-3.4\pm1.0$ & $-1.3\pm1.0$ & $3.7\pm 1.0$ & $101\pm 8$ \\
    3--4 & $-3.5\pm0.9$ & $1.0\pm0.9$ & $3.6\pm 0.9$ & $82\pm 7$ \\
    4--5 & $-2.3\pm1.2$ & $0.1\pm1.2$ & $2.3\pm 1.2$ & $89\pm 15$ \\
    5--6 & $-4.0\pm1.6$ & $0.2\pm1.6$ & $4.0\pm 1.6$ & $89\pm 12$ \\
    6--7 & $-8.5\pm2.2$ & $1.1\pm2.2$ & $8.6\pm 2.2$ & $86\pm 7$ \\
    7--8 & $-2.6\pm3.9$ & $2.5\pm3.9$ & $3.6\pm 3.9$ & $68\pm 31$ \\
    \hline
     2--8 & $-3.8\pm0.7$ & $0.2\pm0.7$ & $3.8\pm0.7$ & $89\pm6$ \\
    \hline
    \end{tabular}
    \label{table:ebins}
\end{table}

\begin{figure}
\centering
\includegraphics[width=0.84\linewidth]{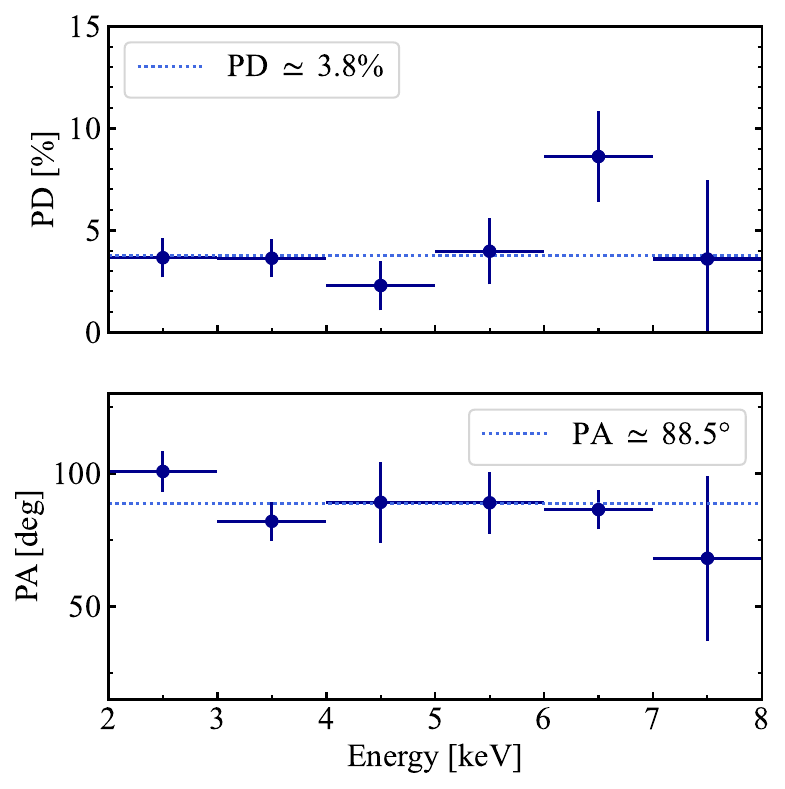}
\caption{Energy dependence of the PD and PA for the combined data set (excluding dip), obtained with the \texttt{pcube} algorithm. } 
\label{fig:energy-dependence}
\end{figure}

\subsection{Polarimetric analysis}

The initial polarimetric analysis  of \smc was carried out using the \texttt{xpbin} tool's \texttt{pcube} algorithm included in the {\sc ixpeobssim} package, which has been implemented according to the formalism by \citet{2015-Kislat}.
Using the unweighted analysis, we compute the normalized Stokes $q=Q/I$ and $u=U/I$ parameters and the PD using the equation PD=$\sqrt{q^2+u^2}$,  and the PA=$\frac{1}{2}\arctan (u/q)$, with the PA measured from north to east counterclockwise on the sky.

\begin{figure}
\centering
\includegraphics[width=0.85\columnwidth]{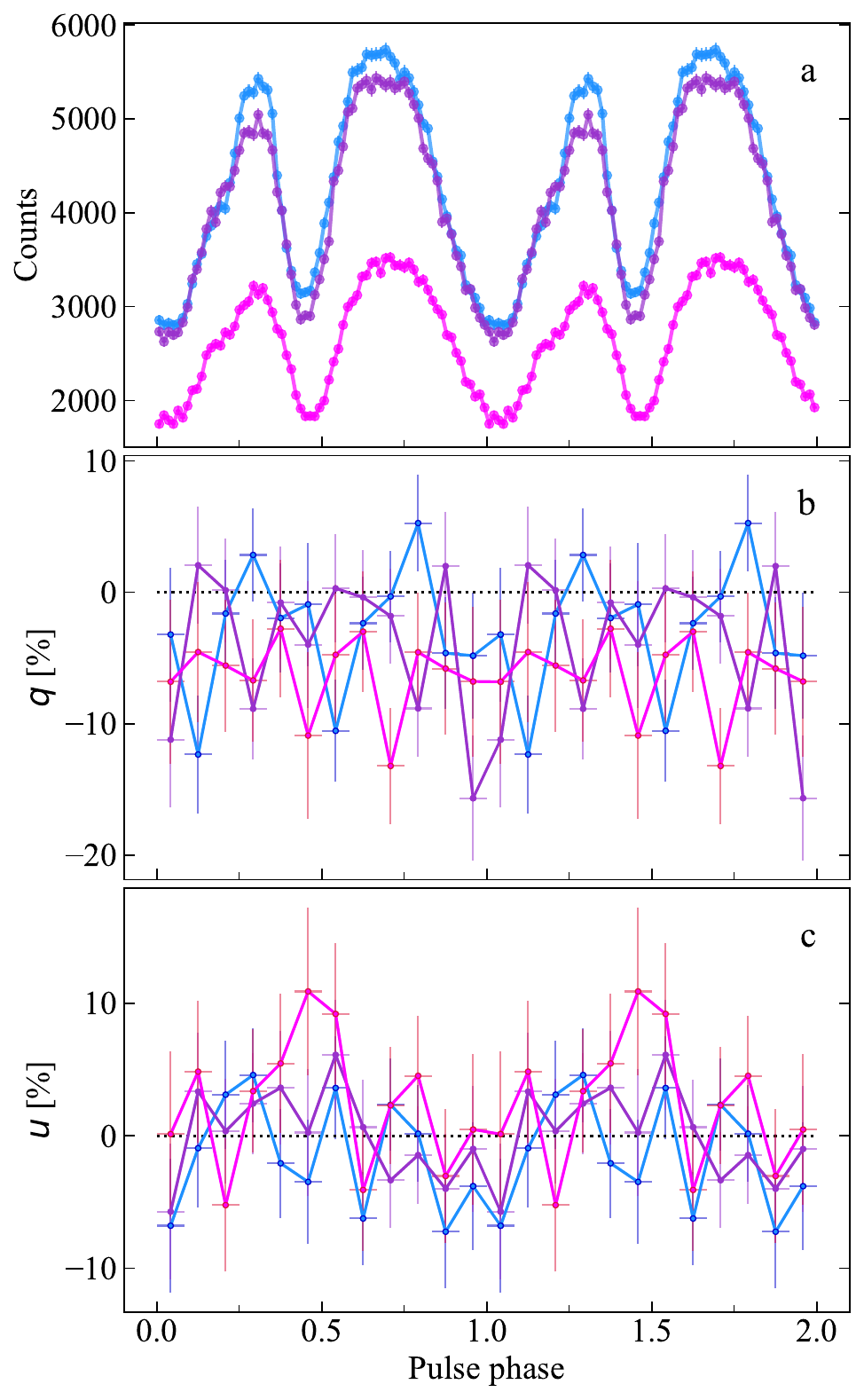}
\caption{Phase dependence of the flux and normalized Stokes parameters. 
Panel (a): pulse profiles during three observations. 
The normalized Stokes parameters $q$ and $u$ are shown in panels (b) and (c), respectively. 
The polarimetric analysis was done using \texttt{pcube} for three DUs combined in the full 2--8 keV energy band and uniform phase-binning.
Observations 1, 2, and 3 are shown in blue, purple, and pink, respectively.}
\label{fig:phase-res-Stokesqu}
\end{figure}


The phase-averaged PD and PA in the full 2--8 keV \ixpe energy range for Obs.~1--3 are given in Table~\ref{table:avebins} and shown in Fig.~\ref{fig:timeres}.
We do not detect polarization during the pre-eclipse dip, with a phase-averaged PD of $0.4\pm3.8\%$ and an unconstrained PA.
The pre-eclipse data is removed from all subsequent analysis.
The PD and PA shows an apparent trend in their behavior over time. The PD appears to be increasing with time, while the PA decreases (in decrements of $\sim10\degr$).

In order to increase the statistics of the polarimetric analysis, we combined the data from all three observations using the following steps. 
First, we created pulse profiles according to the spin period evolution parameters determined for each observation separately (see Table~\ref{table:time-pars}).
The pulse profiles were then cross-correlated to determine the relative phase-shifts between the observations, allowing us to correctly connect the phases.
Finally, each event was phase-tagged and the data from the three observations were added, producing a combined data set (excluding pre-eclipse data).
The phase-averaged PD and PA for the combined set of observations of \smc measured in the entire \ixpe energy band of 2--8 keV are $3.8\pm0.7\%$ and $89\degr\pm 6\degr$, respectively.




Next, we studied the energy dependence of the polarization properties of \smc by performing an energy-resolved polarimetric analysis on the combined data set, dividing the data into six energy bins.
The results are given in Table~\ref{table:ebins} and shown in Fig.~\ref{fig:energy-dependence}.
No significant energy dependence of the polarization properties is detected.
Similarly, we see no energy dependence of the polarimetric properties during the separate observations.



\begin{figure}
\centering
\includegraphics[width=0.85\linewidth]{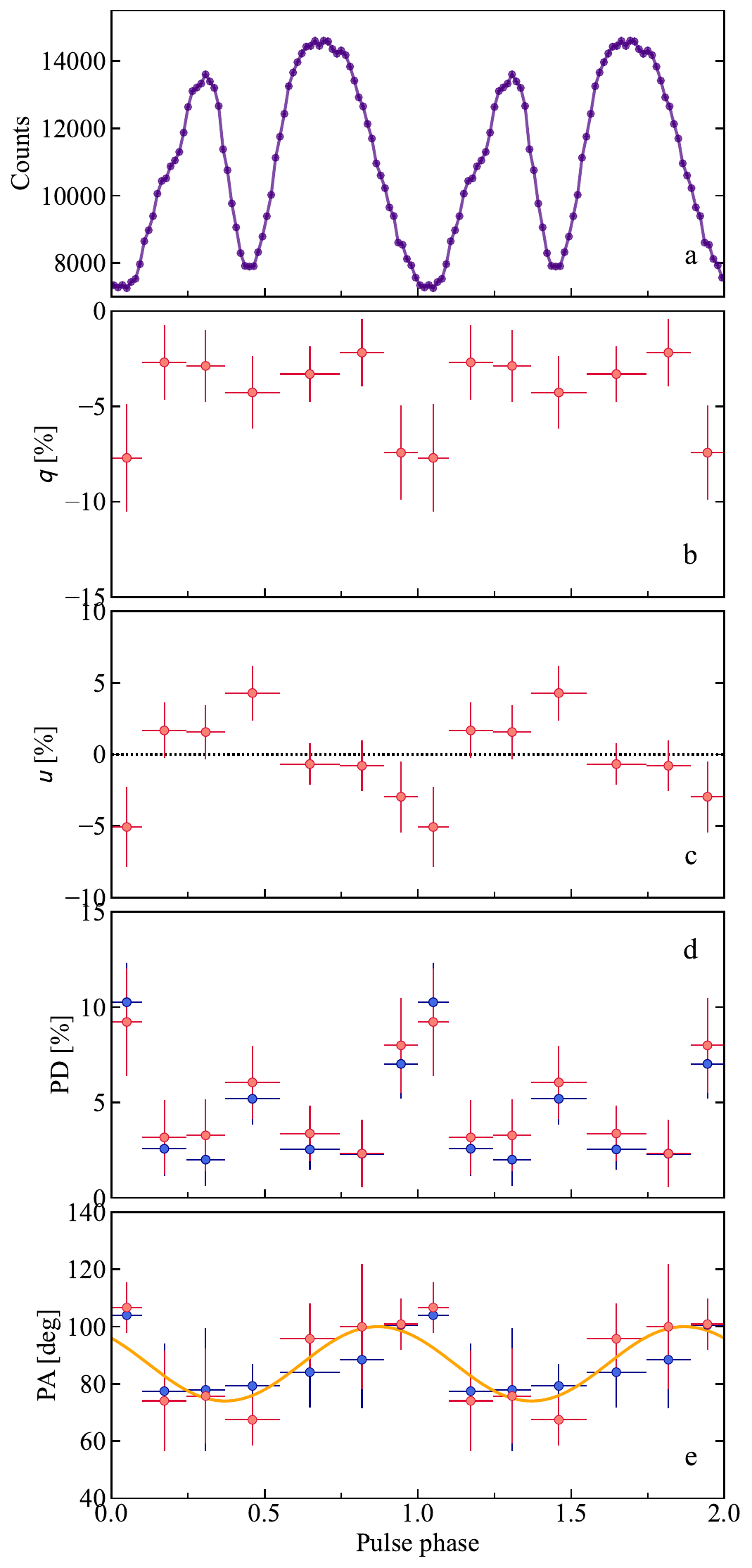}
\caption{
Results from the pulse-phase-resolved analysis of \smc in the 2--8 keV range combining data from all DUs and using non-uniform phase-binning.  
\textit{Panel (a)}: Pulse profile.
\textit{Panels (b)} and \textit{(c)}: Dependence of the Stokes $q$ and $u$ parameters obtained from the \texttt{pcube} algorithm on the pulse phase. 
\textit{Panels (d)} and \textit{(e)}:  PD and PA obtained with \texttt{pcube} and from the phase-resolved spectro-polarimetric analysis using {\sc xspec} (shown by the red and blue symbols, respectively).
The orange curve in panel \textit{(e)} shows the best-fit RVM to the combined data set.
}
\label{fig:phase-res-pcube-xspec}
\end{figure}

Considering the importance of the variations of PD and PA over the pulsar's spin phase, next, we performed a phase-resolved polarimetric analysis of the separate observations by splitting the data into 12 uniform phase bins and using the \texttt{pcube} algorithm to determine the polarimetric properties of each bin.
However, due to poor statistics, we do not detect significant polarization in any phase bin of the separate observations.
We do, however, see an indication of different behavior of the normalized Stokes $q$ and $u$ parameters in the individual observations, as displayed in Fig.~\ref{fig:phase-res-Stokesqu}. 
Similarly, we used the same steps to perform a phase-resolved polarimetric analysis of the combined data set, which indicated an anti-correlation between the PD and flux.
Therefore, we applied a non-uniform phase-binning to the combined data set, with the  phase bins chosen to cover the maxima and minima of the pulse profile.
The results in the 2--8 keV energy range are given in Table~\ref{table:phase-res-PCUBE} and are shown in Fig.~\ref{fig:phase-res-pcube-xspec}.



\begin{table} 
\centering
\caption{Normalized $q$ and $u$ Stokes parameters and PD and PA in different phase bins obtained from the \texttt{pcube} algorithm (combined data set; dip excluded).}
\begin{tabular}{ccccc}
    \hline\hline
    Phase & $q$ & $u$ & PD & PA \\
           & (\%) & (\%) & (\%) & (deg) \\
    \hline
    0.000--0.100 & $-7.7\pm2.8$ & $-5.1\pm2.8$ & $9.2\pm2.8$ & $107\pm9$ \\
    0.100--0.245 & $-2.7\pm1.9$ & $1.7\pm1.9$ & $3.2\pm1.9$ & $74\pm18$ \\
    0.245--0.370 & $-2.9\pm1.9$ & $1.6\pm1.9$ & $3.3\pm1.9$ & $76\pm17$ \\
    0.370--0.550 & $-4.3\pm1.9$ & $4.3\pm1.9$ & $6.1\pm1.9$ & $67\pm9$ \\
    0.550--0.745 & $-3.3\pm1.5$ & $-0.7\pm1.5$ & $3.4\pm1.5$ & $96\pm12$ \\
    0.745--0.890 & $-2.2\pm1.8$ & $-0.8\pm1.8$ & $2.3\pm1.8$ & $100\pm22$ \\
    0.890--1.000 & $-7.4\pm2.5$ & $-3.0\pm2.5$ & $8.0\pm2.5$ & $101\pm9$ \\
    \hline
\end{tabular}
\tablefoot{The uncertainties are given at the 68.3\% (1$\sigma$) confidence level. }
\label{table:phase-res-PCUBE}
\end{table}



\begin{table*}[]
\centering
\caption{Spectral parameters for the best-fit model obtained from the phase-averaged spectro-polarimetric analysis with \textsc{xspec} for observations 1, 2, and 3.}
\begin{tabular}{llccccc}
\hline \hline
Component & Parameter & Unit & Obs. 1 & Obs. 2 & Obs. 3 & Obs.~1--3 \\
\hline
\texttt{tbabs} & $N_{\mathrm{H}}$ & $10^{22}\mathrm{\;cm^{-2}}$ & $0.23\pm0.07$ & $0.31\pm0.07$ & $0.43\pm0.09$ & $0.31\pm0.05$ \\
\texttt{powerlaw} & Photon index &  & $1.00\pm0.02$ & $1.02\pm0.02$ & $1.02\pm0.02$ & $1.01\pm0.01$ \\
\texttt{polconst} & PD & \%      & $3.2\pm0.8$ & $3.0\pm0.9$ & $5.5\pm1.1$ & $3.5\pm0.5$ \\
 & PA & deg     & $97\pm8$ & $90\pm8$ & $80\pm6$ & $89\pm4$ \\
\texttt{constant} & $\mathrm{const_{DU2}}$ &  & $1.019\pm0.005$ & $1.021\pm0.005$ & $1.026\pm0.006$ & $1.021\pm0.003$ \\
 & $\mathrm{const_{DU3}}$ &  & $1.002\pm0.005$ & $1.000\pm0.005$ & $1.023\pm0.006$ & $1.001\pm0.003$ \\
\hline
 & $\mathrm{Flux_{2-8\;keV}}$ & $10^{-10}$\,erg\,s$^{-1}$\,cm$^{-2}$  & $5.10\pm0.07$ & $4.92\pm0.07$ & $4.65\pm0.07$ & -- \\
 & $L_{2-8\rm\,keV}$\tablefootmark{a} & $10^{38}$\,\lumcgs  & 2.3 & 2.2 & 2.1 & -- \\
 & $\chi^2$ (d.o.f.) &  & 1332.1 (1259) & 1240.0 (1262) & 1256.9 (1214) & 3845.3 (3747) \\
\hline  
\end{tabular}
\tablefoot{
The uncertainties are given at the 68.3\% (1$\sigma$) confidence level and were obtained using the \texttt{error} command in \textsc{xspec} with $\Delta\chi^2=1$ for one parameter of interest. 
\tablefoottext{a}{Luminosity is computed using the distance of $d=61$~kpc \citep{2005-Hilditch}.}
}
\label{table:best-fit}
\end{table*}

\begin{figure*}
\centering
\includegraphics[width=0.6\columnwidth]{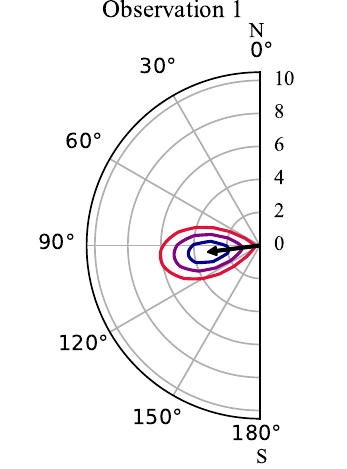}
\includegraphics[width=0.6\columnwidth]{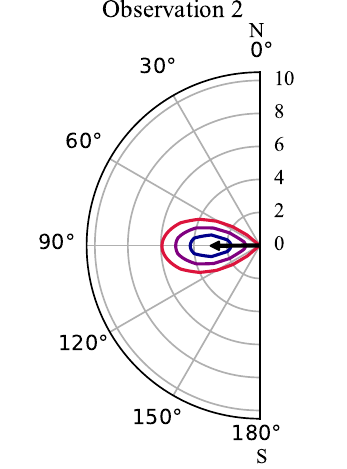}
\includegraphics[width=0.6\columnwidth]{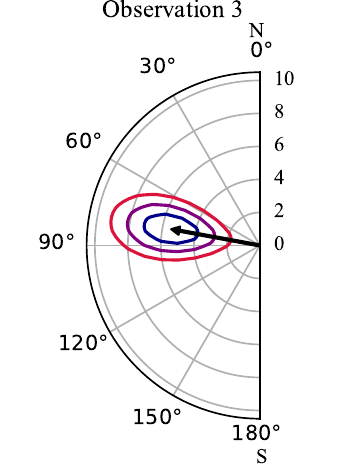}
\caption{Polarization vectors of \smc from the results of the phase-averaged spectro-polarimetric analysis of Obs.~1 (\textit{left}), Obs.~2 (\textit{center}), and Obs.~3 (\textit{right}). Contours at 68.3\%, 95.45\%, and 99.73\% confidence levels calculated for two degrees of freedom are shown in blue, purple, and red, respectively.}
\label{fig:obs-contours}
\end{figure*}

To fully account for the spectral shape and the energy dispersion, we performed a spectro-polarimetric analysis according to the following steps. 
Source Stokes $I$, $Q$, and $U$ spectra were extracted using the \texttt{xpbin} tool's \texttt{PHA1}, \texttt{PHA1Q}, and \texttt{PHA1U} algorithms, which produce a full data set made up of nine spectra, three for each DU.
We fitted all nine spectra  simultaneously with {\sc xspec}. 

The spectral continuum of \smc has been described by a number of different phenomenological models, however, it is usually represented by an absorbed power law with a high-energy cut-off and a soft component, as well as a weak iron line at about 6.4~keV.
Taking into consideration the energy range and resolution of \ixpe, we adopted a simplified model consisting of an absorbed \texttt{powerlaw}, the \texttt{polconst} polarization model (energy-independent PA and PD), as well as a cross-calibration constant accounting for possible discrepancies between the different DUs, where the value for DU1 was fixed at unity.
The final spectral model,
\begin{eqnarray*} 
\texttt{ tbabs$\times$polconst$\times$powerlaw$\times$const},
\label{eq:fit}
\end{eqnarray*}
was applied to both the phase-averaged and phase-resolved data.
The spectral analysis was performed over the full 2--8 keV energy range of \ixpe.
The steppar command in \textsc{xspec} was used to create the confidence contours for the phase-averaged polarization measurements of Obs. 1, 2 and 3, and the resulting contour plots at 68.3\%, 95.45\%, and 99.73\% confidence levels are shown in Fig.~\ref{fig:obs-contours}.
Table~\ref{table:best-fit} lists the spectral parameters for the best-fit models from the results of the phase-averaged spectro-polarimetric analysis for Obs. 1, 2, and 3.
We find significant detection of polarization corresponding to $3.3\sigma$, $3.0\sigma$, and $4.7\sigma$ for Obs. 1, 2, and 3, respectively, determined for two degrees of freedom.

To examine the possibility of any energy-dependence of the polarization properties during the individual observations, the \texttt{polconst} polarization model was replaced with the \texttt{pollin} model (linear energy dependence of the PD and PA) and the \texttt{polpow} model (power law energy dependence of the PD and PA).
Neither model resulted in a significant improvement of the fit.


Considering the similar values of the spectral parameters between all three observations, a joint spectro-polarimetric fit was performed at the next step by fitting the spectra of all three observations simultaneously (with a total of 27 spectra, nine per observation).
The power-law normalization was allowed to vary between the observations.
The results of the phase-averaged spectro-polarimetric analysis of the combined set of spectra can be found in Table~\ref{table:best-fit}.
Similarly to the analysis of the separate observations, the energy-dependence of the polarization properties for the joint fit was tested by replacing the \texttt{polconst} polarization model with the \texttt{pollin} model and \texttt{polpow} model. However, neither model significantly improved the fit.



Next, a phase-resolved spectro-polarimetric analysis was performed by separating the data into seven non-uniform phase-bins.
This was achieved by extracting Stokes $I$, $Q$, and $U$ spectra individually for each phase bin, using the \texttt{PHA1}, \texttt{PHA1Q}, and \texttt{PHA1U} algorithms of the \texttt{xpbin} tool.

\begin{figure*}
\centering
\includegraphics[width=0.2\linewidth]{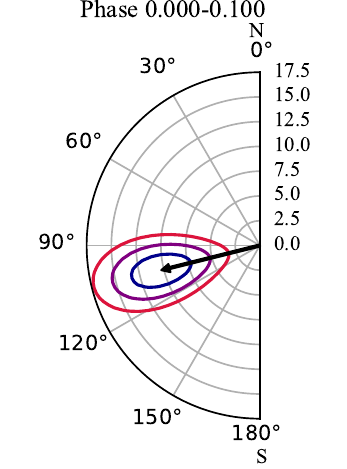}
\includegraphics[width=0.2\linewidth]{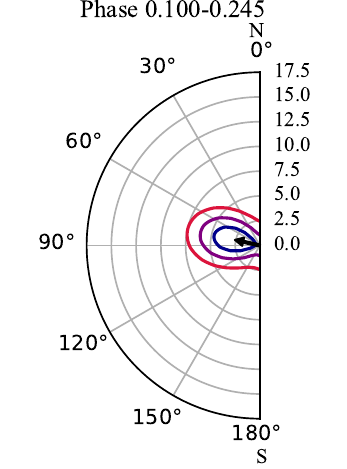}
\includegraphics[width=0.2\linewidth]{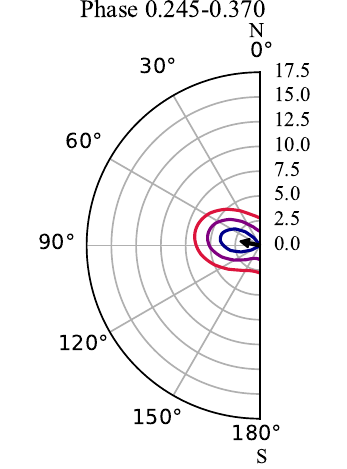}
\includegraphics[width=0.2\linewidth]{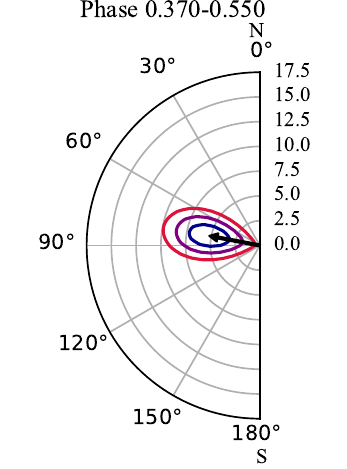}
\includegraphics[width=0.2\linewidth]{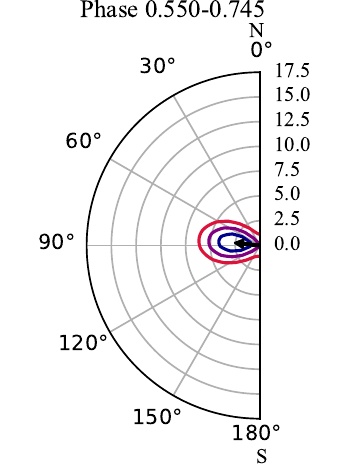}
\includegraphics[width=0.2\linewidth]{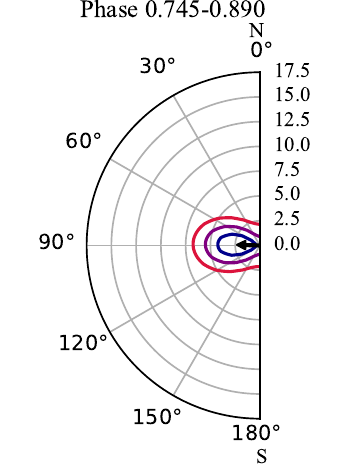}
\includegraphics[width=0.2\linewidth]{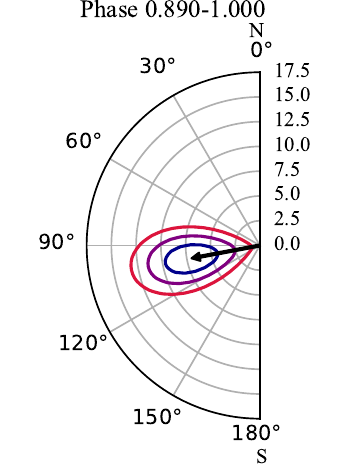}
\caption{Polarization vectors of \smc from the results of the phase-resolved spectro-polarimetric analysis of the combined data set. Contours at 68.3\%, 95.45\%, and 99.73\% confidence levels calculated for two degrees of freedom are shown in blue, purple, and red, respectively.
}
 \label{fig:phase-res-xspec}
\end{figure*}

The phase-resolved $I$, $Q$, and $U$ spectra were fitted with the same model as used for the phase-averaged analysis, and the cross-calibration constants for DU2 and DU3 were set to the values of the phase-averaged analysis (see Table~\ref{table:best-fit}).
The steppar command in \textsc{xspec} was used to create the confidence contours for the phase-resolved polarization measurements, and the resulting contour plots at 68.3\%, 95.45\%, and 99.73\% confidence levels are shown in Fig.~\ref{fig:phase-res-xspec}.
The results of the phase-resolved spectro-polarimetric analysis of the combined data set are summarized in Table~\ref{table:phase-res}.
We find significant detection of polarization in three out of seven phase bins, and marginal detections in the remaining bins.



\begin{table*} 
\centering
\caption{Spectro-polarimetric parameters in different pulse-phase bins for the combined data set obtained with \textsc{xspec}. }
\begin{tabular}{cccccc}
    \hline\hline
    Phase &  $N_{\mathrm{H}}$  & Photon index & PD & PA & $\chi^2$/d.o.f. \\ 
           & ($10^{22}\mathrm{\;cm^{-2}}$) &   & (\%) & (deg) &  \\ 
    \hline
    0.000--0.100 & $0.31$         & $1.05\pm0.02$ & $10.3\pm2.0$ & $104\pm6$ & 2038/2058 \\
    0.100--0.245 & $0.25\pm0.13$  & $0.96\pm0.03$ & $2.6\pm1.4$ & $78\pm17$ & 2679/2621 \\   
    0.245--0.370 & $0.83\pm0.12$  & $1.10\pm0.03$ & $2.0\pm1.4$ & $78\pm22$ & 2699/2657 \\  
    0.370--0.550 & $0.31$         & $1.11\pm 0.01$ & $5.2\pm1.4$ & $79\pm8$ & 2623/2658 \\  
    0.550--0.745 & $0.56\pm0.09$  & $1.06\pm 0.02$ & $2.5\pm1.1$ & $84\pm12$ & 3032/3047 \\   
    0.745--0.890 & $0.38\pm0.11$  & $1.04\pm 0.03$ & $2.3\pm1.3$ & $88\pm17$ & 2743/2759 \\  
    0.890--1.000 & $0.31$         & $1.08\pm0.01$ & $7.0\pm1.8$ & $101\pm8$ & 2230/2247 \\ 
    \hline
    \end{tabular}
\tablefoot{For the first, fourth, and last phase bin, the value of $N_{\rm H}$ was fixed to the value of the phase-averaged spectro-polarimetric analysis for the combined data set. 
The uncertainties computed using the \texttt{error} command 
are given at the 68.3\% (1$\sigma$) confidence level ($\Delta\chi^2=1$ for one parameter of interest). 
} 
\label{table:phase-res}
\end{table*}

\begin{figure*}
\centering
\includegraphics[width=0.85\linewidth]{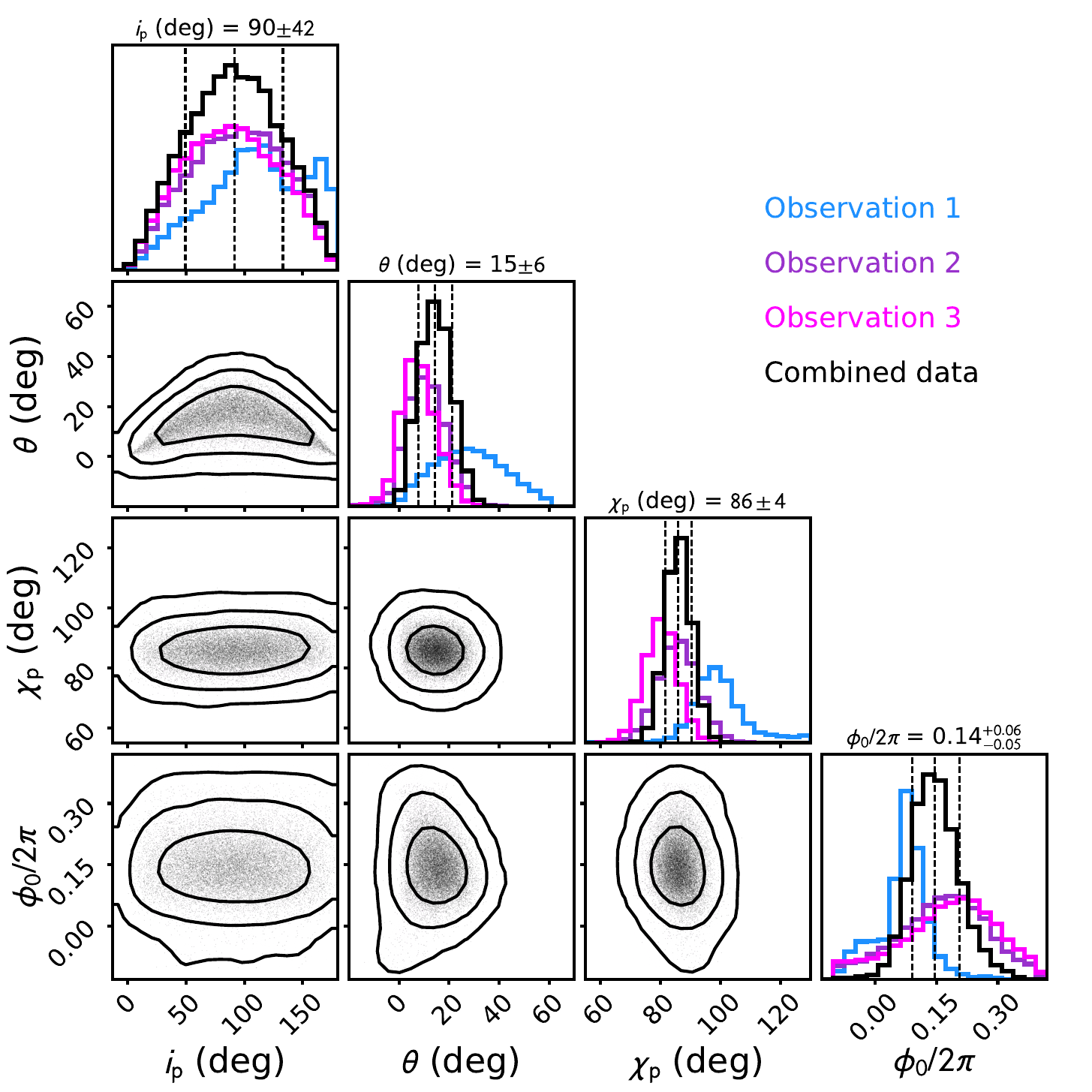}
\caption{Corner plot of the posterior distribution for parameters of the RVM  model fitted directly to the $(q,u)$ values using the likelihood function given by Eq.~\eqref{eq:PA_dist}. 
The two-dimensional contours correspond to 68.3\%, 95.45\% and 99.73\% confidence levels and are shown for the combined data set. 
The histograms show the normalized one-dimensional distributions for a given parameter derived from the posterior samples and are displayed for the individual observations and the combined data set. 
}
\label{fig:rvm}
\end{figure*}

\section{Discussion} 
\label{sec:discussion}

\subsection{Determination of pulsar geometry}

\begin{table}
\caption{Best-fit RVM parameters for the separate observations of \smc, as well as the combined data set.} 
\centering
\begin{tabular}{lcccc}
\hline
\hline 
Parameter  & Obs. 1 & Obs. 2 & Obs. 3 & Obs.~1--3 \\
\hline
$i_{\rm p}$ (deg)        & $98^{+38}_{-41}$ & $100^{+40}_{-45}$ & $88^{+43}_{-43}$ & $91^{+41}_{-42}$ \\
$\theta$ (deg)  & $23^{+14}_{-14}$ & $14^{+8}_{-8}$ &  $6^{+7}_{-6}$ & $13^{+7}_{-6}$  \\
$\chi_{\rm p}$ (deg)    & $98^{+13}_{-6}$ & $86^{+6}_{-7}$ & $80^{+6}_{-5}$ & $87^{+4}_{-4}$ \\
$\phi_0/2\pi$  & $0.07^{+0.05}_{-0.04}$ & $0.14^{+0.08}_{-0.09}$ & $0.18^{+0.17}_{-0.15}$ & $0.12^{+0.08}_{-0.06}$ \\
\hline
\end{tabular}
\label{table:rvm}
\end{table} 

The rotating vector model \citep[RVM;][]{Radhakrishnan69,Meszaros88} can be used to constrain the pulsar geometry.
The geometrical properties of several XRPs observed by \ixpe have already been obtained \citep{2022NatAs...6.1433D,2022ApJ...941L..14T,2023MNRAS.524.2004M,2022ApJ...940...70M,2023A&A...675A..48T,2023A&A...675A..29M,2023A&A...677A..57D,Suleimanov2023,Heyl24,2024MNRAS.tmp.1187Z}. 
If the radiation is assumed to be dominated by ordinary mode (O-mode) photons, the PA is given by equation\,(30) from \citet{Poutanen2020}:
\begin{equation} \label{eq:pa_rvm}
\tan (\mbox{PA}\!-\!\chi_{\rm p})\!=\! \frac{-\sin \theta\ \sin (\phi-\phi_0)}
{\sin i_{\rm p} \cos \theta\!  - \! \cos i_{\rm p} \sin \theta  \cos (\phi\!-\!\phi_0) } ,
\end{equation} 
where $\chi_{\rm p}$ is the position angle (measured from north to east) of the pulsar angular momentum, $i_{\rm p}$ is the inclination of the pulsar spin to the line of sight, $\theta$ comprises the angle between the magnetic dipole and the spin axis, and $\phi_0$ equals the phase when the northern magnetic pole passes in front of the observer. 

If the radiation escapes predominantly in the extraordinary mode (X-mode), the position angle of the pulsar angular momentum is $\chi_{\rm p}\pm90\degr$.
The PA does not depend on the PD of the radiation escaping from the surface of the NS in the no-relativistic RVM.
The polarization plane actually rotates as the radiation travels through the NS magnetosphere, up to the adiabatic radius.
At such a distance, the dipole magnetic field component will dominate, and under these conditions, the RVM is applicable.
Only when the NS is rotating rapidly will general relativistic effects have an effect on the polarization plane \citep{Poutanen2020}.

We can fit the RVM to the measured Stokes $q$ and $u$ parameters, which are normally distributed, as a function of the pulsar phase. 
Because the PA is not normally distributed, we use the probability density function of the PA, $\psi$, from \citet{Naghizadeh1993}:
\begin{equation} \label{eq:PA_dist}
G(\psi) = \frac{1}{\sqrt{\pi}} 
\left\{  \frac{1}{\sqrt{\pi}}  + 
\eta {\rm e}^{\eta^2} 
\left[ 1 + {\rm erf}(\eta) \right]
\right\} {\rm e}^{-p_0^2/2}.
\end{equation}
Here, $p_0=\sqrt{q^2+u^2}/\sigma_{\rm p}$ is the measured PD in units of the error,  $\eta=p_0 \cos[2(\psi-\psi_0)]/\sqrt{2}$, $\psi_0=\frac{1}{2}\arctan(u/q)$ is the central PA obtained from the Stokes parameters, and \mbox{erf} is the error function.

The RVM can be fitted to the pulse-phase dependent $(q,u)$  obtained from \texttt{pcube} using the affine invariant Markov Chain Monte Carlo (MCMC) ensemble sampler {\sc emcee} package of {\sc python} \citep{2013PASP..125..306F} and applying the likelihood function $L= \Pi_i  G(\psi_i)$ with the product taken over all phase bins.  
The RVM was fitted to both the separate observations and the combined set of observations.
The best-fit RVM parameters are given in Table~\ref{table:rvm}.
The covariance plot for the parameters is shown in Fig.~\ref{fig:rvm}.   
The RVM provides an overall good fit to the combined data set and the separate observations.

We see in Fig.~\ref{fig:rvm} that the position angle of the pulsar $\chi_{\rm p}$ shows changes over the course of the observations, decreasing in roughly $10\degr$ decrements. Similarly, the PD also exhibits changes between the individual observations, increasing over time and with decreasing luminosities.
The observations of \smc were carried out during the high-state, corresponding to different super-orbital phases.
The super-orbital variability is generally associated with the precession of the accretion disk, causing periodical obscuration of the central source.
Scattering in the wind of the disk may introduce a polarized component which could lead to variations in the PA pulse phase dependence depending on the super-orbital phase. Similar conclusions have been drawn for other XRPs observed with \ixpe that also display variations in the pulse phase dependence of the PA \citep{2023A&A...677A..57D,2024MNRAS.tmp.1187Z,Poutanen24}.

\subsection{Anti-correlation of luminosity and PD}

\smc is the brightest XRP observed by \ixpe so far.
The super-critical luminosity indicates the presence of an accretion column \citep{1976MNRAS.175..395B}, which is expected to result in high PDs \citep{2021MNRAS.501..109C}.
However, the PD for \smc is relatively low, consistent with other XRPs observed by \ixpe.

\begin{figure}
\centering
\includegraphics[width=0.95\linewidth]{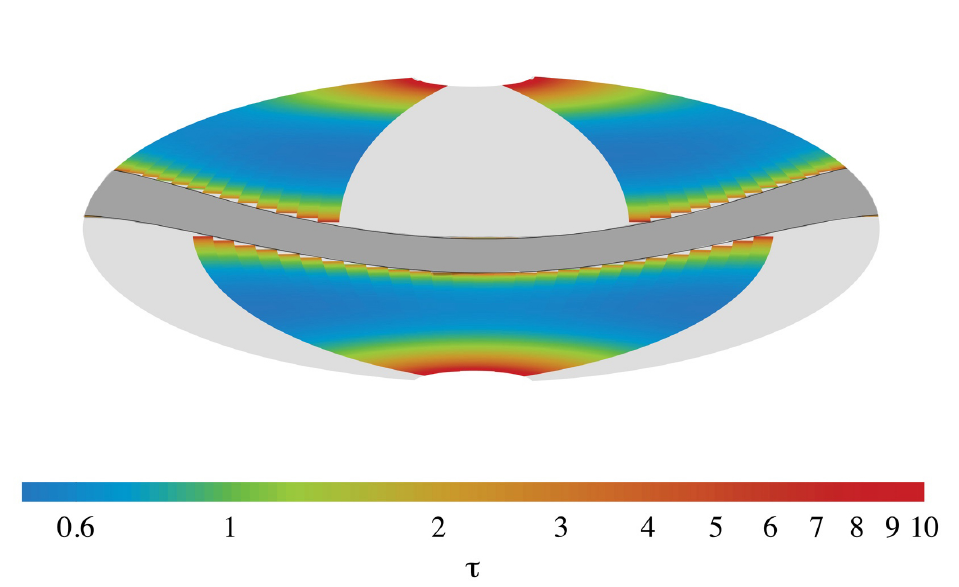}
\caption{Map of the optical thickness $\tau$ (due to the Compton scattering) distribution over the magnetosphere of \smc calculated using the model of \citet{2024MNRAS.530..730M} as viewed from the NS center in Aitoff projection with the $z$-axis aligned along the magnetic dipole. 
The color represents the distribution of the optical thickness.
The gray belt is the region where the accretion disk touches NS magnetosphere and where $\tau$ is infinite. 
The magnetosphere is covered only partially because of the assumed $15\degr$ inclination of magnetic dipole with respect to the disk axis. 
The flow is transparent at azimuthal angle around 0\degr\ in the northern hemisphere and at the angle 180\degr\ in the southern hemisphere, because there is no accretion along these field lines.
The NS spin axis is assumed to be aligned with the disk rotation axis.
We used the following parameters: 
$L=2\times 10^{38}$\,erg\,s$^{-1}$ and 
$R_{\rm m}=5\times 10^{7}$\,cm.
}
\label{fig:sc_env_acc_05_smc_x1}
\end{figure}

The lower degree of linear polarization at higher apparent luminosities  in \smc can be related to the geometry of the emitting region in the super-critical regime of accretion and/or the contribution of the magnetospheric accretion flow.
The appearance of an accretion column at luminosities $\gtrsim 10^{37}$\lumcgs  results in the illumination of the NS surface, and the contribution of reflected X-rays (see, e.g., \citealt{2013ApJ...777..115P,2015MNRAS.452.1601P}) to the apparent photon energy flux and the polarization.
In this case, the PD can be reduced due to the possibility of different contributions of X- and O-modes to the total energy flux. 

At luminosities $\gtrsim 10^{38}$\,\lumcgs, the accretion flow between the inner disk radius and the NS surface can be optically thick for Compton scattering, and therefore scatters a fraction of X-ray photons \citep{2017MNRAS.467.1202M,2019MNRAS.484..687M,2023MNRAS.525.4176B}.
Figure\,\ref{fig:sc_env_acc_05_smc_x1} illustrates the expected distribution of the optical thickness over the magnetospheric surface of the NS in \smc calculated following the model of \citet{2024MNRAS.530..730M}.
The major process responsible for the opacity of the magnetospheric accretion flow at $E>1\,{\rm keV}$ is Compton scattering. 
Because of the magnetic dipole inclination with respect to the rotational axis of the NS, the accretion flow covers only a fraction of the magnetospheric surface.
The optical thickness is expected to be larger (up to $\tau\sim 10$) in regions located close to the plane of the accretion disk and close to the NS surface, because of the relatively high local surface density of the flow.
The minimal optical thickness in the regions of the magnetosphere covered by the accretion flow is $\tau_{\rm min}\approx 0.7$.

For the observed luminosity in \smc ($\sim 2\times 10^{38}$\,\lumcgs) and a NS rotation axis aligned with the disk axis, the magnetospheric flow scatters up to 40\% of X-ray photons depending on the beam pattern of X-ray emission of the NS surface.
For an inclined rotator, the contribution of photons scattered in the magnetosphere to the total flux depends on the observer's viewing angle, but remains close to 40\% on average. 
Because the magnetospheric radius $R_{\rm m}$ is expected to be larger than the adiabatic radius $R_{\rm ad}$ \citep{2016MNRAS.459.3585G,2024Galax..12....6T}
\begin{equation} \label{eq:R_ad}
R_{\rm m}>R_{\rm ad}\simeq
1.2\times 10^7 \left(\frac{B}{10^{12}\,{\rm G}}\right)^{2/5}\!\left(\frac{E}{1\,{\rm keV}}\right)^{1/5}\!\left(\frac{R_{\rm NS}}{10^6\,{\rm cm}}\right)^{6/5}\!\!\mbox{cm},
\end{equation}
the scattering of X-ray photons at the magnetospheric surface can result in depolarization.
The higher the luminosity and optical thickness of the flow, the lower the PD. 

The appearance of an optically thick flow that covers only a fraction of the NS magnetosphere in \smc is aligned with the presence of a pulsating soft blackbody component at energies $<2$\,keV reported in this object \citep[see, e.g.,][]{2002ApJ...579..411P}.
Note, however, that the soft excess in \smc can also be explained by the reflection of X-ray photons off the accretion disk twisted close to the inner radius \citep[see][]{2005ApJ...633.1064H}. 
If this is indeed the case, both the PD and the PA can be affected by X-ray reflection from such a disk.

\section{Summary}
\label{sec:sum}

\smc was observed by \ixpe three separate times around 2023 December 10--20, during the high-state of the super-orbital period at luminosities of $L\sim2\times10^{38}$\,erg\,s$^{-1}$.
The results of the polarimetric analysis of \smc can be summarized as follows:
\begin{enumerate}
\item A significant polarization was detected during all three observations (spectro-polarimetric analysis), with values of the phase-averaged PD and PA corresponding to $3.2\pm 0.8$\% and $97\degr\pm 8\degr$ for Observation 1, $3.0\pm 0.9$\% and $90\degr\pm 8\degr$ for Observation 2, and $5.5\pm 1.1$\% and $80\degr\pm 6\degr$ for Observation 3. There is an indication of a trend in the behavior of the PD and PA over time.
\item There is no evidence for any energy-dependent nature of the polarization properties found in the polarimetric analysis using \texttt{pcube} algorithm or in the spectro-polarimetric analysis. This holds for the phase-averaged analysis, as well as for the phase-resolved analysis.
\item The phase-resolved spectro-polarimetric analysis found significant detection of polarization in three out of seven phase bins, with the PD ranging between $\sim$2\% and $\sim$10\%, and a corresponding range in the PA of $\sim$70\degr\ to $\sim$100\degr. The PD displays an anti-correlation with the flux amplitude.
\item The phase-resolved polarimetric analysis of the individual observations suggests a difference in the behavior of the normalized Stokes $q$ and $u$ that indicate a change in polarization properties over time. Using the RVM to model the PA pulse phase dependence, we determine the pulsar geometry for the separate observations. The position angle of the pulsar rotation axis displays an evolution with super-orbital phase  supporting the idea of a NS and/or accretion disk precession in this source.    
\end{enumerate}




\begin{acknowledgements}
The Imaging X-ray Polarimetry Explorer (IXPE) is a joint US and Italian mission. The US contribution is supported by the National Aeronautics and Space Administration (NASA) and led and managed by its Marshall Space Flight Center (MSFC), with industry partner Ball Aerospace (contract NNM15AA18C).  The Italian contribution is supported by the Italian Space Agency (Agenzia Spaziale Italiana, ASI) through contract ASI-OHBI-2022-13-I.0, agreements ASI-INAF-2022-19-HH.0 and ASI-INFN-2017.13-H0, and its Space Science Data Center (SSDC) with agreements ASI-INAF-2022-14-HH.0 and ASI-INFN 2021-43-HH.0, and by the Istituto Nazionale di Astrofisica (INAF) and the Istituto Nazionale di Fisica Nucleare (INFN) in Italy.  This research used data products provided by the IXPE Team (MSFC, SSDC, INAF, and INFN) and distributed with additional software tools by the High-Energy Astrophysics Science Archive Research Center (HEASARC), at NASA Goddard Space Flight Center (GSFC).

This research has been supported by the Academy of Finland grants 333112 and 349144 (SVF, SST, JP, VL), the German Academic Exchange Service (DAAD) travel grant 57525212 (VD, VFS), UKRI Stephen Hawking fellowship (AAM), and  Deutsche  Forschungsgemeinschaft (DFG) grant WE 1312/59-1 (VFS).
SVF acknowledges the support of the Vilho, Yrjö, and Kalle Väisälä foundation.
The work of RT, GM, FM and PS was partially funded by the Italian Ministry of University and Research (MUR) through grant PRIN 2022LWPEXW. IL was supported by the NASA Postdoctoral Program at the Marshall Space Flight Center, administered by Oak Ridge Associated Universities under contract with NASA. 
\end{acknowledgements}

\bibliographystyle{yahapj}
\bibliography{allbib}

\begin{thebibliography}{}
\providecommand\natexlab[1]{#1}
\providecommand\JournalTitle[1]{#1}

\bibitem[{{Angelini} {et~al.}(1991){Angelini}, {Stella}, \& {White}}]{1991-Angelini}
{Angelini}, L., {Stella}, L., \& {White}, N.~E. 1991, \href{http://dx.doi.org/10.1086/169895}{\JournalTitle{\apj}, 371, 332}

\bibitem[{{Arnaud}(1996)}]{Arn96}
{Arnaud}, K.~A. 1996, in ASP Conf. Ser., Vol. 101, Astronomical Data Analysis Software and Systems V, ed. G.~H. {Jacoby} \& J.~{Barnes} (San Francisco: Astron. Soc. Pac.), 17

\bibitem[{{Baldini} {et~al.}(2021){Baldini}, {Barbanera}, {Bellazzini}, {Bonino}, {Borotto}, {Brez}, {Caporale}, {Cardelli}, {Castellano}, {Ceccanti}, {Citraro}, {Di Lalla}, {Latronico}, {Lucchesi}, {Magazz{\`u}}, {Magazz{\`u}}, {Maldera}, {Manfreda}, {Marengo}, {Marrocchesi}, {Mereu}, {Minuti}, {Mosti}, {Nasimi}, {Nuti}, {Oppedisano}, {Orsini}, {Pesce-Rollins}, {Pinchera}, {Profeti}, {Sgr{\`o}}, {Spandre}, {Tardiola}, {Zanetti}, {Amici}, {Andersson}, {Attin{\`a}}, {Bachetti}, {Baumgartner}, {Brienza}, {Carpentiero}, {Castronuovo}, {Cavalli}, {Cavazzuti}, {Centrone}, {Costa}, {D'Alba}, {D'Amico}, {Del Monte}, {Di Cosimo}, {Di Marco}, {Di Persio}, {Donnarumma}, {Evangelista}, {Fabiani}, {Ferrazzoli}, {Kitaguchi}, {La Monaca}, {Lefevre}, {Loffredo}, {Lorenzi}, {Mangraviti}, {Matt}, {Meilahti}, {Morbidini}, {Muleri}, {Nakano}, {Negri}, {Nenonen}, {O'Dell}, {Perri}, {Piazzolla}, {Pieraccini}, {Pilia}, {Puccetti}, {Ramsey}, {Rankin}, {Ratheesh}, {Rubini}, {Santoli}, {Sarra}, {Scalise}, {Sciortino}, {Soffitta},
  {Tamagawa}, {Tennant}, {Tobia}, {Trois}, {Uchiyama}, {Vimercati}, {Weisskopf}, {Xie}, {Zanetti}, \& {Zhou}}]{2021APh...13302628B}
{Baldini}, L., {Barbanera}, M., {Bellazzini}, R., {et~al.} 2021, \href{http://dx.doi.org/10.1016/j.astropartphys.2021.102628}{\JournalTitle{Astroparticle Physics}, 133, 102628}

\bibitem[{{Basko} \& {Sunyaev}(1976)}]{1976MNRAS.175..395B}
{Basko}, M.~M., \& {Sunyaev}, R.~A. 1976, \href{http://dx.doi.org/10.1093/mnras/175.2.395}{\JournalTitle{\mnras}, 175, 395}

\bibitem[{{Bonnet-Bidaud} \& {van der Klis}(1981)}]{1981-Bonnet-Bidaud}
{Bonnet-Bidaud}, J.~M., \& {van der Klis}, M. 1981, \JournalTitle{\aap}, 97, 134

\bibitem[{{Brice} {et~al.}(2023){Brice}, {Zane}, {Taverna}, {Turolla}, \& {Wu}}]{2023MNRAS.525.4176B}
{Brice}, N., {Zane}, S., {Taverna}, R., {Turolla}, R., \& {Wu}, K. 2023, \href{http://dx.doi.org/10.1093/mnras/stad2391}{\JournalTitle{\mnras}, 525, 4176}

\bibitem[{{Caiazzo} \& {Heyl}(2021)}]{2021MNRAS.501..109C}
{Caiazzo}, I., \& {Heyl}, J. 2021, \href{http://dx.doi.org/10.1093/mnras/staa3428}{\JournalTitle{\mnras}, 501, 109}

\bibitem[{{Clarkson} {et~al.}(2003){Clarkson}, {Charles}, {Coe}, {Laycock}, {Tout}, \& {Wilson}}]{2003-Clarkson}
{Clarkson}, W.~I., {Charles}, P.~A., {Coe}, M.~J., {et~al.} 2003, \href{http://dx.doi.org/10.1046/j.1365-8711.2003.06176.x}{\JournalTitle{\mnras}, 339, 447}

\bibitem[{{Dage} {et~al.}(2019){Dage}, {Clarkson}, {Charles}, {Laycock}, \& {Shih}}]{2019-Dage}
{Dage}, K.~C., {Clarkson}, W.~I., {Charles}, P.~A., {Laycock}, S. G.~T., \& {Shih}, I.~C. 2019, \href{http://dx.doi.org/10.1093/mnras/sty2572}{\JournalTitle{\mnras}, 482, 337}

\bibitem[{{Di Marco} {et~al.}(2022){Di Marco}, {Costa}, {Muleri}, {Soffitta}, {Fabiani}, {La Monaca}, {Rankin}, {Xie}, {Bachetti}, {Baldini}, {Baumgartner}, {Bellazzini}, {Brez}, {Castellano}, {Del Monte}, {Di Lalla}, {Ferrazzoli}, {Latronico}, {Maldera}, {Manfreda}, {O'Dell}, {Perri}, {Pesce-Rollins}, {Puccetti}, {Ramsey}, {Ratheesh}, {Sgr{\`o}}, {Spandre}, {Tennant}, {Tobia}, {Trois}, \& {Weisskopf}}]{Di_Marco_2022}
{Di Marco}, A., {Costa}, E., {Muleri}, F., {et~al.} 2022, \href{http://dx.doi.org/10.3847/1538-3881/ac51c9}{\JournalTitle{\aj}, 163, 170}

\bibitem[{{Di Marco} {et~al.}(2023){Di Marco}, {Soffitta}, {Costa}, {Ferrazzoli}, {La Monaca}, {Rankin}, {Ratheesh}, {Xie}, {Baldini}, {Del Monte}, {Ehlert}, {Fabiani}, {Kim}, {Muleri}, {O'Dell}, {Ramsey}, {Rubini}, {Sgr{\`o}}, {Silvestri}, {Tennant}, \& {Weisskopf}}]{Di_Marco_2023}
{Di Marco}, A., {Soffitta}, P., {Costa}, E., {et~al.} 2023, \href{http://dx.doi.org/10.3847/1538-3881/acba0f}{\JournalTitle{\aj}, 165, 143}

\bibitem[{{Doroshenko} {et~al.}(2022){Doroshenko}, {Poutanen}, {Tsygankov}, {Suleimanov}, {Bachetti}, {Caiazzo}, {Costa}, {Di Marco}, {Heyl}, {La Monaca}, \& et~al.}]{2022NatAs...6.1433D}
{Doroshenko}, V., {Poutanen}, J., {Tsygankov}, S.~S., {et~al.} 2022, \href{http://dx.doi.org/10.1038/s41550-022-01799-5}{\JournalTitle{Nature Astronomy}, 6, 1433}

\bibitem[{{Doroshenko} {et~al.}(2023){Doroshenko}, {Poutanen}, {Heyl}, {Tsygankov}, {Caiazzo}, {Turolla}, {Veledina}, {Weisskopf}, {Forsblom}, {Gonz{\'a}lez-Caniulef}, \& et~al.}]{2023A&A...677A..57D}
{Doroshenko}, V., {Poutanen}, J., {Heyl}, J., {et~al.} 2023, \href{http://dx.doi.org/10.1051/0004-6361/202347088}{\JournalTitle{\aap}, 677, A57}

\bibitem[{{Foreman-Mackey} {et~al.}(2013){Foreman-Mackey}, {Hogg}, {Lang}, \& {Goodman}}]{2013PASP..125..306F}
{Foreman-Mackey}, D., {Hogg}, D.~W., {Lang}, D., \& {Goodman}, J. 2013, \href{http://dx.doi.org/10.1086/670067}{\JournalTitle{\pasp}, 125, 306}

\bibitem[{{Forsblom} {et~al.}(2023){Forsblom}, {Poutanen}, {Tsygankov}, {Bachetti}, {Di Marco}, {Doroshenko}, {Heyl}, {La Monaca}, {Malacaria}, {Marshall}, \& et~al.}]{2023ApJ...947L..20F}
{Forsblom}, S.~V., {Poutanen}, J., {Tsygankov}, S.~S., {et~al.} 2023, \href{http://dx.doi.org/10.3847/2041-8213/acc391}{\JournalTitle{\apjl}, 947, L20}

\bibitem[{{Gehrels} {et~al.}(2004){Gehrels}, {Chincarini}, {Giommi}, {Mason}, {Nousek}, {Wells}, {White}, {Barthelmy}, {Burrows}, {Cominsky}, {Hurley}, {Marshall}, {M{\'e}sz{\'a}ros}, {Roming}, {Angelini}, {Barbier}, {Belloni}, {Campana}, {Caraveo}, {Chester}, {Citterio}, {Cline}, {Cropper}, {Cummings}, {Dean}, {Feigelson}, {Fenimore}, {Frail}, {Fruchter}, {Garmire}, {Gendreau}, {Ghisellini}, {Greiner}, {Hill}, {Hunsberger}, {Krimm}, {Kulkarni}, {Kumar}, {Lebrun}, {Lloyd-Ronning}, {Markwardt}, {Mattson}, {Mushotzky}, {Norris}, {Osborne}, {Paczynski}, {Palmer}, {Park}, {Parsons}, {Paul}, {Rees}, {Reynolds}, {Rhoads}, {Sasseen}, {Schaefer}, {Short}, {Smale}, {Smith}, {Stella}, {Tagliaferri}, {Takahashi}, {Tashiro}, {Townsley}, {Tueller}, {Turner}, {Vietri}, {Voges}, {Ward}, {Willingale}, {Zerbi}, \& {Zhang}}]{Gehrels04}
{Gehrels}, N., {Chincarini}, G., {Giommi}, P., {et~al.} 2004, \href{http://dx.doi.org/10.1086/422091}{\JournalTitle{\apj}, 611, 1005}

\bibitem[{{Gonz{\'a}lez Caniulef} {et~al.}(2016){Gonz{\'a}lez Caniulef}, {Zane}, {Taverna}, {Turolla}, \& {Wu}}]{2016MNRAS.459.3585G}
{Gonz{\'a}lez Caniulef}, D., {Zane}, S., {Taverna}, R., {Turolla}, R., \& {Wu}, K. 2016, \href{http://dx.doi.org/10.1093/mnras/stw804}{\JournalTitle{\mnras}, 459, 3585}

\bibitem[{{Heyl} {et~al.}(2024){Heyl}, {Doroshenko}, {Gonz{\'a}lez-Caniulef}, {Caiazzo}, {Poutanen}, {Mushtukov}, {Tsygankov}, {Kirmizibayrak}, {Bachetti}, {Pavlov}, {Forsblom}, {Malacaria}, {Suleimanov}, {Agudo}, {Antonelli}, {Baldini}, {Baumgartner}, {Bellazzini}, {Bianchi}, {Bongiorno}, {Bonino}, {Brez}, {Bucciantini}, {Capitanio}, {Castellano}, {Cavazzuti}, {Chen}, {Ciprini}, {Costa}, {De Rosa}, {Del Monte}, {Di Gesu}, {Di Lalla}, {Di Marco}, {Donnarumma}, {Dov{\v{c}}iak}, {Ehlert}, {Enoto}, {Evangelista}, {Fabiani}, {Ferrazzoli}, {Garcia}, {Gunji}, {Hayashida}, {Iwakiri}, {Jorstad}, {Kaaret}, {Karas}, {Kislat}, {Kitaguchi}, {Kolodziejczak}, {Krawczynski}, {La Monaca}, {Latronico}, {Liodakis}, {Maldera}, {Manfreda}, {Marin}, {Marinucci}, {Marscher}, {Marshall}, {Massaro}, {Matt}, {Mitsuishi}, {Mizuno}, {Muleri}, {Negro}, {Ng}, {O'Dell}, {Omodei}, {Oppedisano}, {Papitto}, {Peirson}, {Perri}, {Pesce-Rollins}, {Petrucci}, {Pilia}, {Possenti}, {Puccetti}, {Ramsey}, {Rankin}, {Ratheesh}, {Roberts}, {Romani},
  {Sgr{\`o}}, {Slane}, {Soffitta}, {Spandre}, {Swartz}, {Tamagawa}, {Tavecchio}, {Taverna}, {Tawara}, {Tennant}, {Thomas}, {Tombesi}, {Trois}, {Turolla}, {Vink}, {Weisskopf}, {Wu}, {Xie}, \& {Zane}}]{Heyl24}
{Heyl}, J., {Doroshenko}, V., {Gonz{\'a}lez-Caniulef}, D., {et~al.} 2024, \href{http://dx.doi.org/10.48550/arXiv.2311.03667}{\JournalTitle{Nature Astronomy, in press}, arXiv:2311.03667}

\bibitem[{{Hickox} \& {Vrtilek}(2005)}]{2005ApJ...633.1064H}
{Hickox}, R.~C., \& {Vrtilek}, S.~D. 2005, \href{http://dx.doi.org/10.1086/491596}{\JournalTitle{\apj}, 633, 1064}

\bibitem[{{Hilditch} {et~al.}(2005){Hilditch}, {Howarth}, \& {Harries}}]{2005-Hilditch}
{Hilditch}, R.~W., {Howarth}, I.~D., \& {Harries}, T.~J. 2005, \href{http://dx.doi.org/10.1111/j.1365-2966.2005.08653.x}{\JournalTitle{\mnras}, 357, 304}

\bibitem[{{Hu} {et~al.}(2013){Hu}, {Chou}, {Yang}, \& {Su}}]{2013-Hu}
{Hu}, C.-P., {Chou}, Y., {Yang}, T.-C., \& {Su}, Y.-H. 2013, \href{http://dx.doi.org/10.1088/0004-637X/773/1/58}{\JournalTitle{\apj}, 773, 58}

\bibitem[{{Hu} {et~al.}(2019){Hu}, {Mihara}, {Sugizaki}, {Ueda}, \& {Enoto}}]{2019-Hu-orb}
{Hu}, C.-P., {Mihara}, T., {Sugizaki}, M., {Ueda}, Y., \& {Enoto}, T. 2019, \href{http://dx.doi.org/10.3847/1538-4357/ab48e4}{\JournalTitle{\apj}, 885, 123}

\bibitem[{{Kaaret} {et~al.}(2017){Kaaret}, {Feng}, \& {Roberts}}]{2017ARA&A..55..303K}
{Kaaret}, P., {Feng}, H., \& {Roberts}, T.~P. 2017, \href{http://dx.doi.org/10.1146/annurev-astro-091916-055259}{\JournalTitle{\araa}, 55, 303}

\bibitem[{{Kislat} {et~al.}(2015){Kislat}, {Clark}, {Beilicke}, \& {Krawczynski}}]{2015-Kislat}
{Kislat}, F., {Clark}, B., {Beilicke}, M., \& {Krawczynski}, H. 2015, \href{http://dx.doi.org/10.1016/j.astropartphys.2015.02.007}{\JournalTitle{Astroparticle Physics}, 68, 45}

\bibitem[{{Leong} {et~al.}(1971){Leong}, {Kellogg}, {Gursky}, {Tananbaum}, \& {Giacconi}}]{1971-Leong}
{Leong}, C., {Kellogg}, E., {Gursky}, H., {Tananbaum}, H., \& {Giacconi}, R. 1971, \href{http://dx.doi.org/10.1086/180842}{\JournalTitle{\apjl}, 170, L67}

\bibitem[{{Lucke} {et~al.}(1976){Lucke}, {Yentis}, {Friedman}, {Fritz}, \& {Shulman}}]{1976-Lucke}
{Lucke}, R., {Yentis}, D., {Friedman}, H., {Fritz}, G., \& {Shulman}, S. 1976, \href{http://dx.doi.org/10.1086/182125}{\JournalTitle{\apjl}, 206, L25}

\bibitem[{{Malacaria} {et~al.}(2023){Malacaria}, {Heyl}, {Doroshenko}, {Tsygankov}, {Poutanen}, {Forsblom}, {Capitanio}, {Di Marco}, {Du}, {Ducci}, \& et~al.}]{2023A&A...675A..29M}
{Malacaria}, C., {Heyl}, J., {Doroshenko}, V., {et~al.} 2023, \href{http://dx.doi.org/10.1051/0004-6361/202346581}{\JournalTitle{\aap}, 675, A29}

\bibitem[{{Marshall} {et~al.}(2022){Marshall}, {Ng}, {Rogantini}, {Heyl}, {Tsygankov}, {Poutanen}, {Costa}, {Zane}, {Malacaria}, {Agudo}, {Antonelli}, {Bachetti}, {Baldini}, {Baumgartner}, {Bellazzini}, {Bianchi}, {Bongiorno}, {Bonino}, {Brez}, {Bucciantini}, {Capitanio}, {Castellano}, {Cavazzuti}, {Ciprini}, {De Rosa}, {Del Monte}, {Di Gesu}, {Di Lalla}, {Di Marco}, {Donnarumma}, {Doroshenko}, {Dov{\v{c}}iak}, {Ehlert}, {Enoto}, {Evangelista}, {Fabiani}, {Ferrazzoli}, {Garcia}, {Gunji}, {Hayashida}, {Iwakiri}, {Jorstad}, {Karas}, {Kitaguchi}, {Kolodziejczak}, {Krawczynski}, {Monaca}, {Latronico}, {Liodakis}, {Maldera}, {Manfreda}, {Marin}, {Marinucci}, {Marscher}, {Matt}, {Mitsuishi}, {Mizuno}, {Muleri}, {Ng}, {O'Dell}, {Omodei}, {Oppedisano}, {Papitto}, {Pavlov}, {Peirson}, {Perri}, {Pesce-Rollins}, {Petrucci}, {Pilia}, {Possenti}, {Puccetti}, {Ramsey}, {Rankin}, {Ratheesh}, {Romani}, {Sgr{\`o}}, {Slane}, {Soffitta}, {Spandre}, {Tamagawa}, {Tavecchio}, {Taverna}, {Tawara}, {Tennant}, {Thomas}, {Tombesi},
  {Trois}, {Turolla}, {Vink}, {Weisskopf}, {Wu}, {Xie}, {IXPE Collaboration}, {Schulz}, \& {Chakrabarty}}]{2022ApJ...940...70M}
{Marshall}, H.~L., {Ng}, M., {Rogantini}, D., {et~al.} 2022, \href{http://dx.doi.org/10.3847/1538-4357/ac98c2}{\JournalTitle{\apj}, 940, 70}

\bibitem[{{Meszaros} {et~al.}(1988){Meszaros}, {Novick}, {Szentgyorgyi}, {Chanan}, \& {Weisskopf}}]{Meszaros88}
{Meszaros}, P., {Novick}, R., {Szentgyorgyi}, A., {Chanan}, G.~A., \& {Weisskopf}, M.~C. 1988, \href{http://dx.doi.org/10.1086/165962}{\JournalTitle{\apj}, 324, 1056}

\bibitem[{{Moon} {et~al.}(2003){Moon}, {Eikenberry}, \& {Wasserman}}]{2003-Moon}
{Moon}, D.-S., {Eikenberry}, S.~S., \& {Wasserman}, I.~M. 2003, \href{http://dx.doi.org/10.1086/367782}{\JournalTitle{\apjl}, 582, L91}

\bibitem[{{Mushtukov} \& {Tsygankov}(2024)}]{MushtukovTsygankov2024}
{Mushtukov}, A., \& {Tsygankov}, S. 2024, \href{http://dx.doi.org/10.1007/978-981-19-6960-7_104}{in Handbook of X-ray and Gamma-ray Astrophysics, ed. C.~{Bambi} \& A.~{Santangelo}} (Singapore: Springer), 4105

\bibitem[{{Mushtukov} {et~al.}(2019){Mushtukov}, {Ingram}, {Middleton}, {Nagirner}, \& {van der Klis}}]{2019MNRAS.484..687M}
{Mushtukov}, A.~A., {Ingram}, A., {Middleton}, M., {Nagirner}, D.~I., \& {van der Klis}, M. 2019, \href{http://dx.doi.org/10.1093/mnras/sty3525}{\JournalTitle{\mnras}, 484, 687}

\bibitem[{{Mushtukov} {et~al.}(2024){Mushtukov}, {Ingram}, {Suleimanov}, {DiLullo}, {Middleton}, {Tsygankov}, {van der Klis}, \& {Portegies Zwart}}]{2024MNRAS.530..730M}
{Mushtukov}, A.~A., {Ingram}, A., {Suleimanov}, V.~F., {et~al.} 2024, \href{http://dx.doi.org/10.1093/mnras/stae781}{\JournalTitle{\mnras}, 530, 730}

\bibitem[{{Mushtukov} {et~al.}(2017){Mushtukov}, {Suleimanov}, {Tsygankov}, \& {Ingram}}]{2017MNRAS.467.1202M}
{Mushtukov}, A.~A., {Suleimanov}, V.~F., {Tsygankov}, S.~S., \& {Ingram}, A. 2017, \href{http://dx.doi.org/10.1093/mnras/stx141}{\JournalTitle{\mnras}, 467, 1202}

\bibitem[{{Mushtukov} {et~al.}(2023){Mushtukov}, {Tsygankov}, {Poutanen}, {Doroshenko}, {Salganik}, {Costa}, {Di Marco}, {Heyl}, {La Monaca}, {Lutovinov}, \& et~al.}]{2023MNRAS.524.2004M}
{Mushtukov}, A.~A., {Tsygankov}, S.~S., {Poutanen}, J., {et~al.} 2023, \href{http://dx.doi.org/10.1093/mnras/stad1961}{\JournalTitle{\mnras}, 524, 2004}

\bibitem[{{Naghizadeh-Khouei} \& {Clarke}(1993)}]{Naghizadeh1993}
{Naghizadeh-Khouei}, J., \& {Clarke}, D. 1993, \JournalTitle{\aap}, 274, 968

\bibitem[{{Paul} {et~al.}(2002){Paul}, {Nagase}, {Endo}, {Dotani}, {Yokogawa}, \& {Nishiuchi}}]{2002ApJ...579..411P}
{Paul}, B., {Nagase}, F., {Endo}, T., {et~al.} 2002, \href{http://dx.doi.org/10.1086/342701}{\JournalTitle{\apj}, 579, 411}

\bibitem[{{Postnov} {et~al.}(2015){Postnov}, {Gornostaev}, {Klochkov}, {Laplace}, {Lukin}, \& {Shakura}}]{2015MNRAS.452.1601P}
{Postnov}, K.~A., {Gornostaev}, M.~I., {Klochkov}, D., {et~al.} 2015, \href{http://dx.doi.org/10.1093/mnras/stv1393}{\JournalTitle{\mnras}, 452, 1601}

\bibitem[{{Poutanen}(2020)}]{Poutanen2020}
{Poutanen}, J. 2020, \href{http://dx.doi.org/10.1051/0004-6361/202038689}{\JournalTitle{\aap}, 641, A166}

\bibitem[{{Poutanen} {et~al.}(2013){Poutanen}, {Mushtukov}, {Suleimanov}, {Tsygankov}, {Nagirner}, {Doroshenko}, \& {Lutovinov}}]{2013ApJ...777..115P}
{Poutanen}, J., {Mushtukov}, A.~A., {Suleimanov}, V.~F., {et~al.} 2013, \href{http://dx.doi.org/10.1088/0004-637X/777/2/115}{\JournalTitle{\apj}, 777, 115}

\bibitem[{{Poutanen} {et~al.}(2024){Poutanen}, {Tsygankov}, {Doroshenko}, {Forsblom}, {Jenke}, {Kaaret}, {Berdyugin}, {Blinov}, {Kravtsov}, {Liodakis}, {Tzouvanou}, {Di Marco}, {Heyl}, {La Monaca}, {Mushtukov}, {Pavlov}, {Salganik}, {Veledina}, {Weisskopf}, {Zane}, {Loktev}, {Suleimanov}, {Wilson-Hodge}, {Berdyugina}, {Kagitani}, {Piirola}, {Sakanoi}, {Agudo}, {Antonelli}, {Bachetti}, {Baldini}, {Baumgartner}, {Bellazzini}, {Bianchi}, {Bongiorno}, {Bonino}, {Brez}, {Bucciantini}, {Capitanio}, {Castellano}, {Cavazzuti}, {Chen}, {Ciprini}, {Costa}, {De Rosa}, {Del Monte}, {Di Gesu}, {Di Lalla}, {Donnarumma}, {Dovciak}, {Ehlert}, {Enoto}, {Evangelista}, {Fabiani}, {Ferrazzoli}, {Garcia}, {Gunji}, {Hayashida}, {Iwakiri}, {Jorstad}, {Karas}, {Kislat}, {Kitaguchi}, {Kolodziejczak}, {Latronico}, {Maldera}, {Manfreda}, {Marin}, {Marinucci}, {Marscher}, {Marshall}, {Massaro}, {Matt}, {Mitsuishi}, {Mizuno}, {Muleri}, {Negro}, {Ng}, {O'Dell}, {Omodei}, {Oppedisano}, {Papitto}, {Peirson}, {Perri}, {Pesce-Rollins},
  {Petrucci}, {Pilia}, {Possenti}, {Puccetti}, {Ramsey}, {Rankin}, {Ratheesh}, {Roberts}, {Romani}, {Sgro}, {Slane}, {Soffitta}, {Spandre}, {Swartz}, {Tamagawa}, {Tavecchio}, {Taverna}, {Tawara}, {Tennant}, {Thomas}, {Tombesi}, {Trois}, {Turolla}, {Vink}, {Wu}, \& {Xie}}]{Poutanen24}
{Poutanen}, J., {Tsygankov}, S.~S., {Doroshenko}, V., {et~al.} 2024, \href{http://dx.doi.org/10.48550/arXiv.2405.08107}{\JournalTitle{\aap, submitted}, arXiv:2405.08107}

\bibitem[{{Price} {et~al.}(1971){Price}, {Groves}, {Rodrigues}, {Seward}, {Swift}, \& {Toor}}]{1971-Price}
{Price}, R.~E., {Groves}, D.~J., {Rodrigues}, R.~M., {et~al.} 1971, \href{http://dx.doi.org/10.1086/180773}{\JournalTitle{\apjl}, 168, L7}

\bibitem[{{Radhakrishnan} \& {Cooke}(1969)}]{Radhakrishnan69}
{Radhakrishnan}, V., \& {Cooke}, D.~J. 1969, \JournalTitle{\aplett}, 3, 225

\bibitem[{{Rai} {et~al.}(2018){Rai}, {Pradhan}, \& {Paul}}]{2018-Rai}
{Rai}, B., {Pradhan}, P., \& {Paul}, B.~C. 2018, \href{http://dx.doi.org/10.1088/1674-4527/18/12/148}{\JournalTitle{Research in Astronomy and Astrophysics}, 18, 148}

\bibitem[{{Raichur} \& {Paul}(2010)}]{2010-Raichur-orb}
{Raichur}, H., \& {Paul}, B. 2010, \href{http://dx.doi.org/10.1111/j.1365-2966.2009.15778.x}{\JournalTitle{\mnras}, 401, 1532}

\bibitem[{{Reig}(2011)}]{2011-Reig}
{Reig}, P. 2011, \href{http://dx.doi.org/10.1007/s10509-010-0575-8}{\JournalTitle{\apss}, 332, 1}

\bibitem[{{Schreier} {et~al.}(1972){Schreier}, {Giacconi}, {Gursky}, {Kellogg}, \& {Tananbaum}}]{1972-Schreier}
{Schreier}, E., {Giacconi}, R., {Gursky}, H., {Kellogg}, E., \& {Tananbaum}, H. 1972, \href{http://dx.doi.org/10.1086/181086}{\JournalTitle{\apjl}, 178, L71}

\bibitem[{{Soffitta} {et~al.}(2021){Soffitta}, {Baldini}, {Bellazzini}, {Costa}, {Latronico}, {Muleri}, {Del Monte}, {Fabiani}, {Minuti}, {Pinchera}, {Sgro'}, {Spandre}, {Trois}, {Amici}, {Andersson}, {Attina'}, {Bachetti}, {Barbanera}, {Borotto}, {Brez}, {Brienza}, {Caporale}, {Cardelli}, {Carpentiero}, {Castellano}, {Castronuovo}, {Cavalli}, {Cavazzuti}, {Ceccanti}, {Centrone}, {Ciprini}, {Citraro}, {D'Amico}, {D'Alba}, {Di Cosimo}, {Di Lalla}, {Di Marco}, {Di Persio}, {Donnarumma}, {Evangelista}, {Ferrazzoli}, {Hayato}, {Kitaguchi}, {La Monaca}, {Lefevre}, {Loffredo}, {Lorenzi}, {Lucchesi}, {Magazzu}, {Maldera}, {Manfreda}, {Mangraviti}, {Marengo}, {Matt}, {Mereu}, {Morbidini}, {Mosti}, {Nakano}, {Nasimi}, {Negri}, {Nenonen}, {Nuti}, {Orsini}, {Perri}, {Pesce-Rollins}, {Piazzolla}, {Pilia}, {Profeti}, {Puccetti}, {Rankin}, {Ratheesh}, {Rubini}, {Santoli}, {Sarra}, {Scalise}, {Sciortino}, {Tamagawa}, {Tardiola}, {Tobia}, {Vimercati}, \& {Xie}}]{2021AJ....162..208S}
{Soffitta}, P., {Baldini}, L., {Bellazzini}, R., {et~al.} 2021, \href{http://dx.doi.org/10.3847/1538-3881/ac19b0}{\JournalTitle{\aj}, 162, 208}

\bibitem[{{Suleimanov} {et~al.}(2023){Suleimanov}, {Forsblom}, {Tsygankov}, {Poutanen}, {Doroshenko}, {Doroshenko}, {Capitanio}, {Di Marco}, {Gonz{\'a}lez-Caniulef}, {Heyl}, {La Monaca}, {Lutovinov}, {Molkov}, {Malacaria}, {Mushtukov}, {Shtykovsky}, {Agudo}, {Antonelli}, {Bachetti}, {Baldini}, {Baumgartner}, {Bellazzini}, {Bianchi}, {Bongiorno}, {Bonino}, {Brez}, {Bucciantini}, {Castellano}, {Cavazzuti}, {Chen}, {Ciprini}, {Costa}, {De Rosa}, {Del Monte}, {Di Gesu}, {Di Lalla}, {Donnarumma}, {Dov{\v{c}}iak}, {Ehlert}, {Enoto}, {Evangelista}, {Fabiani}, {Ferrazzoli}, {Garcia}, {Gunji}, {Hayashida}, {Iwakiri}, {Jorstad}, {Kaaret}, {Karas}, {Kislat}, {Kitaguchi}, {Kolodziejczak}, {Krawczynski}, {Latronico}, {Liodakis}, {Maldera}, {Manfreda}, {Marin}, {Marinucci}, {Marscher}, {Marshall}, {Massaro}, {Matt}, {Mitsuishi}, {Mizuno}, {Muleri}, {Negro}, {Ng}, {O'Dell}, {Omodei}, {Oppedisano}, {Papitto}, {Pavlov}, {Peirson}, {Perri}, {Pesce-Rollins}, {Petrucci}, {Pilia}, {Possenti}, {Puccetti}, {Ramsey}, {Rankin},
  {Ratheesh}, {Roberts}, {Romani}, {Sgr{\`o}}, {Slane}, {Soffitta}, {Spandre}, {Swartz}, {Tamagawa}, {Tavecchio}, {Taverna}, {Tawara}, {Tennant}, {Thomas}, {Tombesi}, {Trois}, {Turolla}, {Vink}, {Weisskopf}, {Wu}, {Xie}, \& {Zane}}]{Suleimanov2023}
{Suleimanov}, V.~F., {Forsblom}, S.~V., {Tsygankov}, S.~S., {et~al.} 2023, \href{http://dx.doi.org/10.1051/0004-6361/202346994}{\JournalTitle{\aap}, 678, A119}

\bibitem[{{Taverna} \& {Turolla}(2024)}]{2024Galax..12....6T}
{Taverna}, R., \& {Turolla}, R. 2024, \href{http://dx.doi.org/10.3390/galaxies12010006}{\JournalTitle{Galaxies}, 12, 6}

\bibitem[{{Trowbridge} {et~al.}(2007){Trowbridge}, {Nowak}, \& {Wilms}}]{2007-Trowbridge}
{Trowbridge}, S., {Nowak}, M.~A., \& {Wilms}, J. 2007, \href{http://dx.doi.org/10.1086/522075}{\JournalTitle{\apj}, 670, 624}

\bibitem[{{Tsygankov} {et~al.}(2022){Tsygankov}, {Doroshenko}, {Poutanen}, {Heyl}, {Mushtukov}, {Caiazzo}, {Di Marco}, {Forsblom}, {Gonz{\'a}lez-Caniulef}, {Klawin}, {La Monaca}, {Malacaria}, {Marshall}, {Muleri}, {Ng}, {Suleimanov}, {Sunyaev}, {Turolla}, {Agudo}, {Antonelli}, {Bachetti}, {Baldini}, {Baumgartner}, {Bellazzini}, {Bianchi}, {Bongiorno}, {Bonino}, {Brez}, {Bucciantini}, {Capitanio}, {Castellano}, {Cavazzuti}, {Ciprini}, {Costa}, {De Rosa}, {Del Monte}, {Di Gesu}, {Di Lalla}, {Donnarumma}, {Dov{\v{c}}iak}, {Ehlert}, {Enoto}, {Evangelista}, {Fabiani}, {Ferrazzoli}, {Garcia}, {Gunji}, {Hayashida}, {Iwakiri}, {Jorstad}, {Karas}, {Kitaguchi}, {Kolodziejczak}, {Krawczynski}, {Latronico}, {Liodakis}, {Maldera}, {Manfreda}, {Marin}, {Marinucci}, {Marscher}, {Matt}, {Mitsuishi}, {Mizuno}, {Ng}, {O'Dell}, {Omodei}, {Oppedisano}, {Papitto}, {Pavlov}, {Peirson}, {Perri}, {Pesce-Rollins}, {Petrucci}, {Pilia}, {Possenti}, {Puccetti}, {Ramsey}, {Rankin}, {Ratheesh}, {Romani}, {Sgr{\`o}}, {Slane}, {Soffitta},
  {Spandre}, {Tamagawa}, {Tavecchio}, {Taverna}, {Tawara}, {Tennant}, {Thomas}, {Tombesi}, {Trois}, {Vink}, {Weisskopf}, {Wu}, {Xie}, {Zane}, \& {IXPE Collaboration}}]{2022ApJ...941L..14T}
{Tsygankov}, S.~S., {Doroshenko}, V., {Poutanen}, J., {et~al.} 2022, \href{http://dx.doi.org/10.3847/2041-8213/aca486}{\JournalTitle{\apjl}, 941, L14}

\bibitem[{{Tsygankov} {et~al.}(2023){Tsygankov}, {Doroshenko}, {Mushtukov}, {Poutanen}, {Di Marco}, {Heyl}, {La Monaca}, {Forsblom}, {Malacaria}, {Marshall}, \& et~al.}]{2023A&A...675A..48T}
{Tsygankov}, S.~S., {Doroshenko}, V., {Mushtukov}, A.~A., {et~al.} 2023, \href{http://dx.doi.org/10.1051/0004-6361/202346134}{\JournalTitle{\aap}, 675, A48}

\bibitem[{{van der Meer} {et~al.}(2007){van der Meer}, {Kaper}, {van Kerkwijk}, {Heemskerk}, \& {van den Heuvel}}]{2007-vanderMeer}
{van der Meer}, A., {Kaper}, L., {van Kerkwijk}, M.~H., {Heemskerk}, M.~H.~M., \& {van den Heuvel}, E.~P.~J. 2007, \href{http://dx.doi.org/10.1051/0004-6361:20066025}{\JournalTitle{\aap}, 473, 523}

\bibitem[{{Weisskopf} {et~al.}(2022){Weisskopf}, {Soffitta}, {Baldini}, {Ramsey}, {O'Dell}, {Romani}, {Matt}, {Deininger}, {Baumgartner}, {Bellazzini}, {Costa}, {Kolodziejczak}, {Latronico}, {Marshall}, {Muleri}, {Bongiorno}, {Tennant}, {Bucciantini}, {Dovciak}, {Marin}, {Marscher}, {Poutanen}, {Slane}, {Turolla}, {Kalinowski}, {Di Marco}, {Fabiani}, {Minuti}, {La Monaca}, {Pinchera}, {Rankin}, {Sgro'}, {Trois}, {Xie}, {Alexander}, {Allen}, {Amici}, {Andersen}, {Antonelli}, {Antoniak}, {Attina'}, {Barbanera}, {Bachetti}, {Baggett}, {Bladt}, {Brez}, {Bonino}, {Boree}, {Borotto}, {Breeding}, {Brienza}, {Bygott}, {Caporale}, {Cardelli}, {Carpentiero}, {Castellano}, {Castronuovo}, {Cavalli}, {Cavazzuti}, {Ceccanti}, {Centrone}, {Citraro}, {D' Amico}, {D'Alba}, {Di Gesu}, {Del Monte}, {Dietz}, {Di Lalla}, {Di Persio}, {Dolan}, {Donnarumma}, {Evangelista}, {Ferrant}, {Ferrazzoli}, {Ferrie}, {Footdale}, {Forsyth}, {Foster}, {Garelick}, {Gunji}, {Gurnee}, {Head}, {Hibbard}, {Johnson}, {Kelly}, {Kilaru}, {Lefevre},
  {Le Roy}, {Loffredo}, {Lorenzi}, {Lucchesi}, {Maddox}, {Magazzu}, {Maldera}, {Manfreda}, {Mangraviti}, {Marengo}, {Marrocchesi}, {Massaro}, {Mauger}, {McCracken}, {McEachen}, {Mize}, {Mereu}, {Mitchell}, {Mitsuishi}, {Morbidini}, {Mosti}, {Nasimi}, {Negri}, {Negro}, {Nguyen}, {Nitschke}, {Nuti}, {Onizuka}, {Oppedisano}, {Orsini}, {Osborne}, {Pacheco}, {Paggi}, {Painter}, {Pavelitz}, {Pentz}, {Piazzolla}, {Perri}, {Pesce-Rollins}, {Peterson}, {Pilia}, {Profeti}, {Puccetti}, {Ranganathan}, {Ratheesh}, {Reedy}, {Root}, {Rubini}, {Ruswick}, {Sanchez}, {Sarra}, {Santoli}, {Scalise}, {Sciortino}, {Schroeder}, {Seek}, {Sosdian}, {Spandre}, {Speegle}, {Tamagawa}, {Tardiola}, {Tobia}, {Thomas}, {Valerie}, {Vimercati}, {Walden}, {Weddendorf}, {Wedmore}, {Welch}, {Zanetti}, \& {Zanetti}}]{Weisskopf2022}
{Weisskopf}, M.~C., {Soffitta}, P., {Baldini}, L., {et~al.} 2022, \href{http://dx.doi.org/10.1117/1.JATIS.8.2.026002}{\JournalTitle{JATIS}, 8, 026002}

\bibitem[{{Wojdowski} {et~al.}(1998){Wojdowski}, {Clark}, {Levine}, {Woo}, \& {Zhang}}]{1998-Wojdowski}
{Wojdowski}, P., {Clark}, G.~W., {Levine}, A.~M., {Woo}, J.~W., \& {Zhang}, S.~N. 1998, \href{http://dx.doi.org/10.1086/305893}{\JournalTitle{\apj}, 502, 253}

\bibitem[{{Zhao} {et~al.}(2024){Zhao}, {Li}, {Tao}, {Feng}, {Zhang}, {Walter}, {Ge}, {Tong}, {Ji}, {Zhang}, {Qu}, {Huang}, {Ma}, {Zhang}, {Yin}, {Yin}, {Ma}, {Zhao}, {Li}, {Yang}, {Liu}, {Yu}, {Huang}, {Li}, {Li}, {Xiao}, \& {Zhao}}]{2024MNRAS.tmp.1187Z}
{Zhao}, Q.~C., {Li}, H.~C., {Tao}, L., {et~al.} 2024, \href{http://dx.doi.org/10.1093/mnras/stae1173}{\JournalTitle{\mnras}}, \href{http://arxiv.org/abs/2405.00509}{{\sffamily arXiv:2405.00509 [astro-ph.HE]}}

\end{thebibliography}
\end{document}